\documentclass[a4paper,11pt]{article}
\pdfoutput=1 % if your are submitting a pdflatex (i.e. if you have
\usepackage{jinstpub} % for details on the use of the package, please
\usepackage{lineno}
\usepackage{multirow}
\usepackage{comment}
\usepackage{graphicx}
\title{\boldmath Development of a novel, windowless, amorphous selenium based photodetector for use in liquid noble detectors}

\author[a]{M.~Rooks \note{Corresponding author.},}
\author[b]{S.~Abbaszadeh,}
\author[a]{J.~Asaadi,}
\author[c]{M.~Febbraro,}
\author[a,2]{R.W.~Gladen\note{also at Fermi National Accelerator Laboratory.}}
\author[d]{E.~Gramellini,}
\author[b]{K.~Hellier,}
\author[d]{F.~Maria Blaszczyk}
\author[a]{A.D.~McDonald }

\affiliation[a]{University of Texas at Arlington,\\Physics Department Arlington, TX 76019, USA}
\affiliation[b]{University of California, \\Santa Cruz
Electrical and Engineering Department,
Santa Cruz Institute for Particle Physics}
\affiliation[c]{Oak Ridge National Laboratory,\\ Physics Division Oak Ridge, TN 37831, USA}
\affiliation[d]{Fermi National Accelerator Laboratory,\\Batavia, IL USA}

\emailAdd{michael.rooks@mavs.uta.edu }

\abstract{Detection of the vacuum ultraviolet (VUV) scintillation light produced by liquid noble elements is a central challenge in order to fully exploit the available timing, topological, and calorimetric information in detectors leveraging these media. In this paper, we  characterize a novel, windowless amorphous selenium based photodetector with direct sensitivity to VUV light. We present here the manufacturing and experimental setup used to operate this detector at low transport electric fields (2.7-5.2 V/$\mu$m) and across a wide range of temperatures (77K-290K). This work shows that the first proof-of-principle device windowless amorphous selenium is robust under cryogenic conditions, responsive to VUV light at cryogenic temperatures, and preserves argon purity. These findings motivate a continued exploration of amorphous selenium devices for simultaneous detection of scintillation light and ionization charge in noble element detectors.
}

\begin{document}
\maketitle
\flushbottom

%%%%%%%%%%%%%%%%%%%%%%%%%%%%%%%%%%%%%%%%%%%%%
%  The Introduction here
%%%%%%%%%%%%%%%%%%%%%%%%%%%%%%%%%%%%%%%%%%%%%
\section{Introduction}\label{sec:intro}

% https://www.mdpi.com/2410-390X/5/1/4/htm 
%============================================================
%%%%  https://www.nature.com/articles/s41598-020-73437-x

The application and ubiquity of noble liquid detectors in the fields of high energy physics \cite{ARNEODO199795,Acciarri_2017,Acciarri_2020,Anderson:2012vc,Zhang_2011,WILLIS1974221,Abi_2020,DUNE:2020cqd}, medical imaging \cite{adelphi,Hernandez:2020fpm,GRIGNON2007142,CHEPEL1997427}, and rare event searches \cite{Acciarri_2010,AMAUDRUZ20191,Collaboration_2009,Aprile_2012,Akerib_2014,Anton_2019,_lvarez_2013} is due to the many attractive properties these media provide. Charged particles traversing noble liquids deposit energy in the form of scintillation light and the ionization charge. Depending on the application, an experiment may choose to apply an external electric field and collect ionization electrons. Given the anti-correlation between the collected ionization charge and the light yield, this comes with a loss in the overall detected scintillation light.

The collection of the scintillation light is a central tool in noble element detectors as it provides a number of useful experimental handles. Firstly, the scintillation light provides a prompt signal (commonly referred to as $t_0$) which allows to record an accurate time associated with the activity observed. This plays a central role in Time Projection Chambers (TPC's) \cite{doi:10.1063/1.2994775} which collect both charge and light as it allows the inference of the position of the event along the drift dimension from the difference between  $t_0$  and the time the charge signal is registered. Secondly, the combination of the amount of light and charge collected provides a robust estimate of energy deposited in the noble element detector \cite{instruments5010013}. Thirdly, techniques in pulse shape discrimination based on the scintillation light allow to distinguish recoils due to electrons from recoils due to nuclear interactions \cite{Wahl_2014,Akerib_2018}. This provides a powerful tool in rare event searches (such as dark matter applications) to separate signal from background.

Two of the most common liquid nobles, argon (Ar) and xenon (Xe), have very good scintillation light yields with excellent optical transmission properties. %This allows for the size of the detectors using them to become quite large while preserving a high flux of photons observed at the photosensor.
Thus, even detectors as big as several meter cubed preserve a high flux of photons observed at the photosensor. %One key challenge is both of these elements emit their scintillation light in the vacuum ultraviolet (VUV) regime. The typical wavelengths for liquid argon (LAr) and liquid xenon (LXe) are 128 nm and 178 nm respectively. 
One key challenge is that both elements scintillate in the vacuum ultraviolet (VUV). The typical wavelengths are 128 nm for liquid argon (LAr) and 178 nm for liquid xenon (LXe). Common photosensors used to detect low levels of light, e.g. Multi-Pixel Photon Counters (MPPCs), Silicon photomultipliers (SiPMs), and photo multiplier tubes (PMTs), are largely insensitive to this wavelength of light due to their construction and fabrication. More recently, devices custom made to be more sensitive to VUV wavelengths have started to emerge \cite{Jamil_2018,8824427,Pershing_2022}, albeit with relatively low efficiencies, reaching at most 15-20\%. 
A standard solution to the mismatch between photosensors' readout and VUV scintillation light is to deploy a wavelength shifting (WLS) material that absorbs the VUV light and re-emits it via fluorescence at a much longer wavelength (typically in the `blue' wavelength).  The past years have seen substantial R\&D in the field of wavelength shifters and their application to liquid noble detectors \cite{instruments5010004}. Two of the most common WLS materials include 1,1,4,4-tetraphenyl-1,3-butadiene (TPB) and  polyethylene naphthalate (PEN). Despite their ubiquity in application, these WLS materials have a number of drawbacks including their deterioration due to environmental effects \cite{Abraham_2021,Jones_2013}, a complicated delayed emission time \cite{Segreto_2015,Araujo2022}, and a relatively low efficiency for the observation of the re-emmitted photon \cite{Benson2018}.

The difficulties associated with the detection of the VUV photons has inspired research into alternative materials which could potentially be sensitive directly to VUV light. In this paper, we explore an amorphous selenium (aSe) based detector. Ample literature on aSe based direct conversion active matrix flat panel imagers (AMFPI) \cite{Wronski2008-zf} and digital breast tomosynthesis \cite{Zhao2008-tf} has taken place in the field of X-ray imaging. The recently developed ability to perform single-photon-counting (SPC) X-ray experiments using CMOS technology \cite{9377635} makes this material an attractive candidate to explore for different applications. The optical absorption properties of aSe \cite{1951} suggest that the material has excellent efficiency for converting the VUV photons into electron/hole pairs at shallow depths (nm), thus overcoming potential depth-dependent effects observed for X-rays \cite{Shiva1}. Moreover, the transport properties of aSe suggest that with sufficiently small distances between the electrodes, the overall mobilities and lifetimes of the charge carriers should be sufficiently high to be viable for low photon flux applications. \\

This paper explores the viability of a windowless aSe based device for collecting UV light in liquid noble element detectors. As such, the response of the device is characterized as a function of temperature in the range relevant for noble element detectors using UV light. The initial exploration is done at relatively low applied electric fields ($\leq 5$~V/$\mu$m), where SPC is not expected because of the limited charge yield. Future work is planned to explore significantly higher electric fields where the holes in aSe undergo impact ionization and thus liberate additional electron-hole pairs. This process has been shown to cause amplification of the initial signal and can result in avalanche gain \cite{pmid25735277}. The first commercial device utilizing impact ionization in aSe, referred to as high-gain avalanche rushing photoconductor (HARP) tubes \cite{TANIOKA2009S15} were initially commercialized in the late 1980s for the broadcast industry. More recently, novel designs in the electrodes has shown that avalanche multiplication with sensitivity down to SPC levels is possible \cite{pmid23298070}. Thus, the thrust of this work is to perform a characterization of a simple, but novel, aSe based photon detector with its deployment in a cryogenic environment to understand the feasibility and limitations of this device. 

Section \ref{sec:setup} describes the aSe device and the testing apparati used to characterize the boards behavior as a function of temperature. Section \ref{sec:Results} describes the observations and behavior of the aSe based detector. Finally, Section \ref{sec:conclusions} offers some closing thoughts and conclusions.

%%%%%%%%%%%%%%%%%%%%%%%%%%%%%%%%%%%%%%%%%%%%%
%  The Experimental Setup here
%%%%%%%%%%%%%%%%%%%%%%%%%%%%%%%%%%%%%%%%%%%%%
\section{Experimental Setup}\label{sec:setup}

In this section, we describe the aSe device under test. Section \ref{sec:aSeBoards} describes the devices fabrication and characterization. Section \ref{sec:CryoStand} presents the experimental apparatus used to test the aSe boards at cryogenic temperatures and under UV light exposure. Finally, Section \ref{sec:Electronics} describes the custom readout electronics and high voltage supply needed to collect both holes and electrons at various electric fields.

%%%%%%%%%%%%%%%%%%%%%%%%%%%%%%%%%%%%%%%%%%%%%
%  The Amorphous Selenium Boards here
%%%%%%%%%%%%%%%%%%%%%%%%%%%%%%%%%%%%%%%%%%%%%
\subsection{Amorphous Selenium Boards}\label{sec:aSeBoards}

\begin{figure}[htb]
    \centering
    \includegraphics[width=0.45\textwidth]{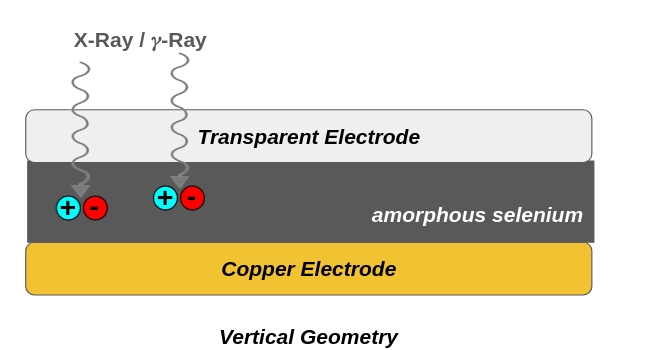}
    \includegraphics[width=0.45\textwidth]{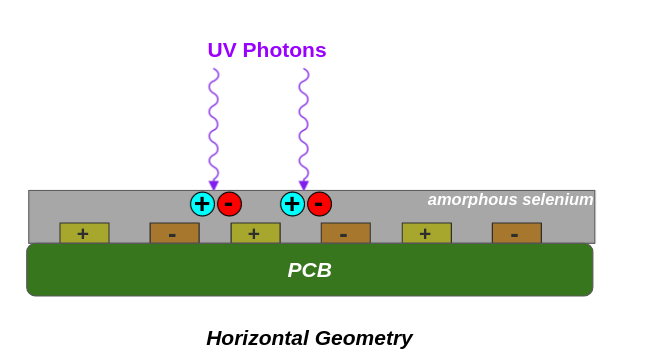}
    \label{fig:VerticalAndHoriztonalGeom}
    \caption{Schematic of the various geometries which an aSe device may be used. Left: ``Vertical Geometry'' which utilizes an electrode on the top most layer which is transparent to the radiation to be detected. Right: ``Horizontal Geometry'' which uses interdigitated electrodes to achieve a horizontal electric field in the aSe and resolves the problem of most electrodes being non-transparent to VUV light.}
\end{figure}

A typical aSe device, as has been used for x-ray and gamma-ray detection \cite{Shiva1}, uses a geometric layout which can be described as a ``vertical geometry''. This geometry has the amorphous selenium sandwiched between two horizontal electrodes, as shown schematically on the left of Figure \ref{fig:VerticalAndHoriztonalGeom}. The electrodes provide an electric field needed to achieve transport of the charge carriers created when a photon interacts with the selenium. The vertical geometry can be used in x-ray and gamma-ray applications because the electrodes are largely transparent to these photons and thus provides a simplified fabrication process. However, for use with UV light this configuration is unfavorable since even a thin amount of material typically used as an upper electrode (e.g. ITO, gold, copper, etc) will result in a large fraction of all the UV light being absorbed. To circumvent this problem and allow for feasibility testing of aSe, we consider a ``horizontal geometry'' such that in Figure \ref{fig:VerticalAndHoriztonalGeom}. 

This geometry consists of a bare printed circuit board (PCB) constructed with interdigitated electrodes to provide the electric field needed to achieve transport of the charge carriers. This configuration thus creates a ``windowless'' device where the selenium is thermally evaporated directly onto the board. The selenium is thus exposed directly to the UV source. This device, as will be shown in this paper, represents a low cost, simple to manufacture, and scalable solution to a large area VUV sensitive photosensor. The PCB manufacturing process for areas as large as 2000 cm$^2$ is commercially ready and low cost \cite{jlcpcb}, the process of uniform and repeatable thermal evaporation techniques over these areas is well demonstrated \cite{9321339}, and the ability to scale together large area tiles into one uniform collection plane is commonly done in experiments \cite{Dwyer_2018,Cantini_2014}. Moreover, the ability for the device to respond to VUV light using this windowless approach simplifies the characterization and testing of the device. A study of the electric field present within the amorphous selenium given the interdigitated electrodes used in the device tested here is presented in the Appendix~\ref{app:EFieldModeling}. This study shows that for the device tested here, the electric field is uniform both across the electrodes as well as throughout the selenium and follows the geometric properties one would intuit. 

\begin{figure}[htb]
    \centering
    \includegraphics[width=0.24\textwidth]{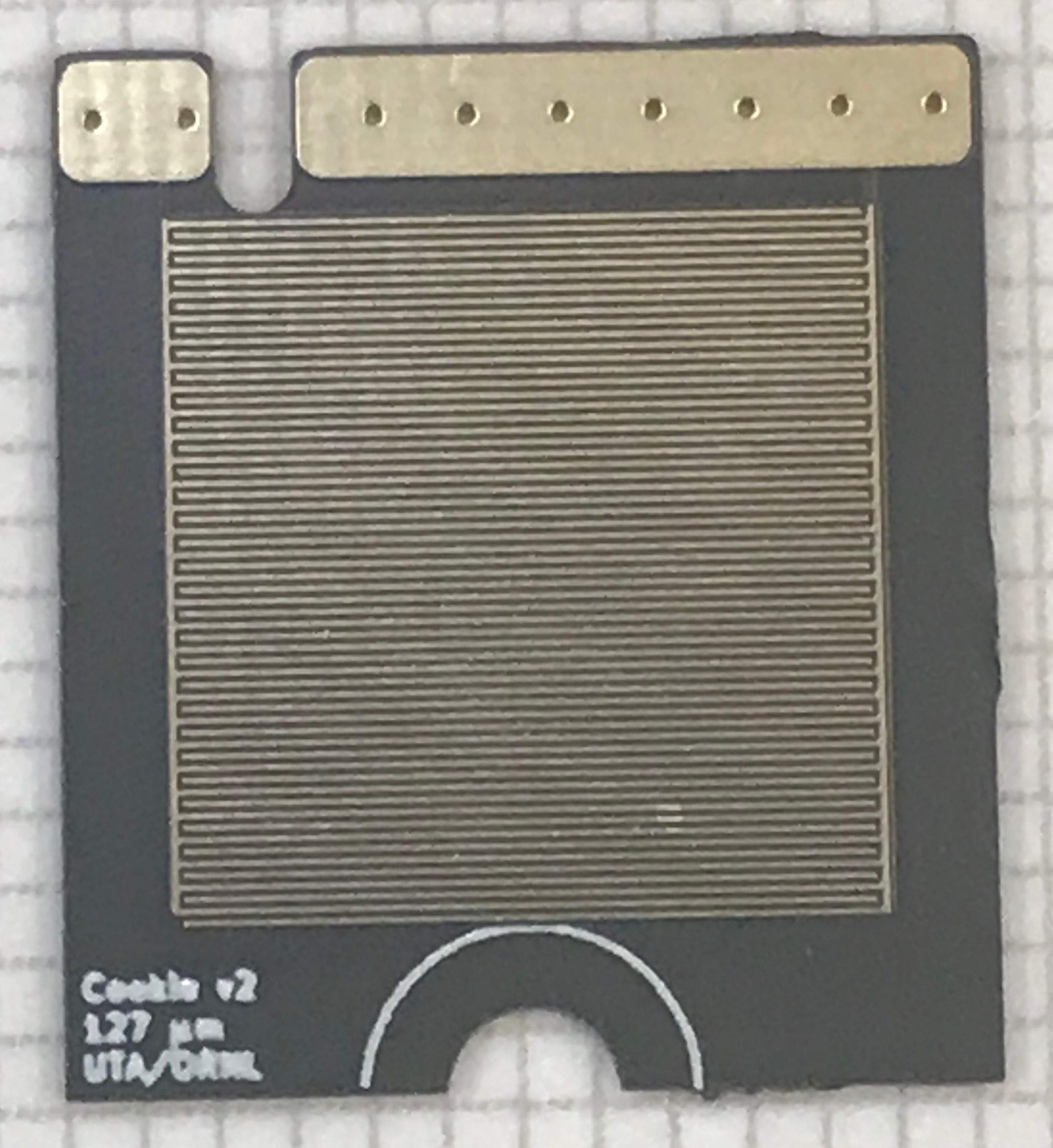}
    \includegraphics[width=0.33\textwidth]{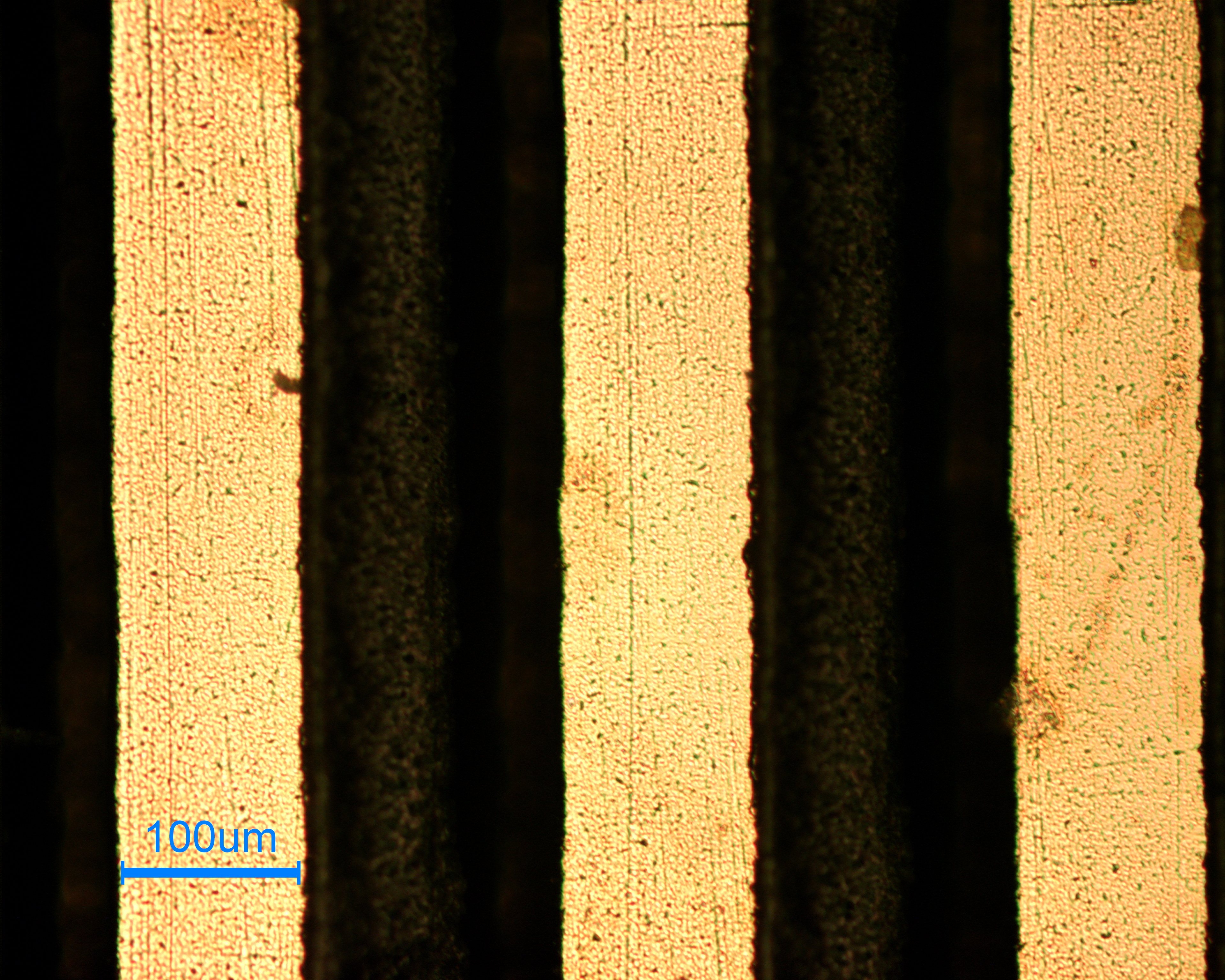}
    \includegraphics[width=0.33\textwidth]{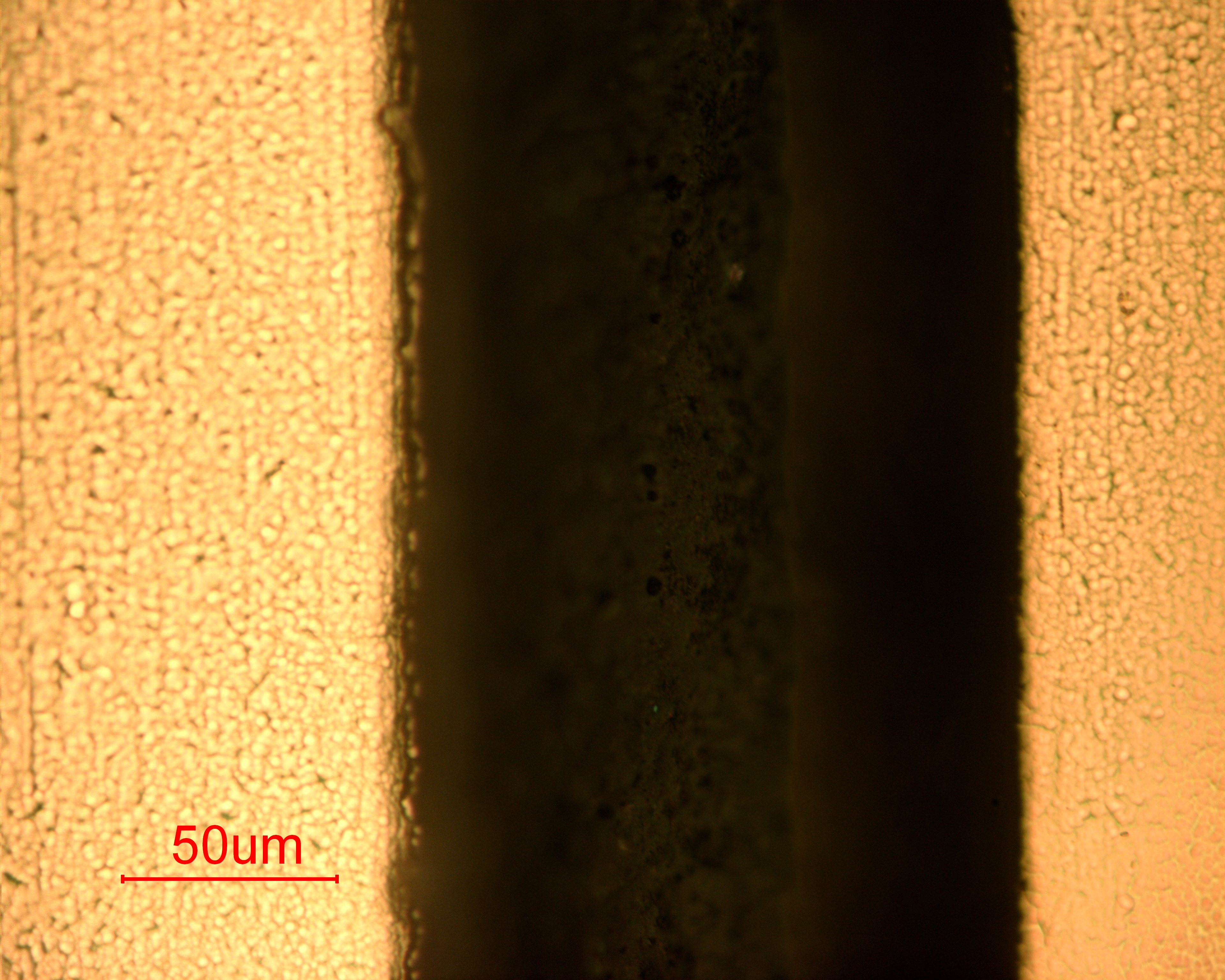}
    \caption{Images of the interdigitated boards used during the experiments described below at various magnifications. The board is FR4-Standard Tg 130-140C and the electrodes are Electroless nickel immersion gold  following the Restriction of Hazardous Substances guidance (ENIG-RoHS).}
    \label{fig:barecookie}
\end{figure}

The horizontal geometry does present some design challenges. The most readily available commercial spacing between PCB produced interdigitated electrodes is limited by the PCB manufacturing process. This results in a limit to the electric field (in units of Volts/micrometer) which can be applied in such a configuration. For the experiment presented here, a small commercial board of 20mm$\times$22.5mm was produced with the smallest electrode spacing of $\sim127 \mu$m from a low-cost commercial vendor \cite{jlcpcb}. The board is shown in Figure \ref{fig:barecookie} before the addition of selenium. 
In order to obtain avalanche gain in the aSe, it is necessary to apply higher fields to the prototypes than the case presented here. Follow-up work on these results will utilize a high density PCB manufacturing process to explore trace separations down to $25\mu$m. However, as the applied field increases and the sensitivity to lower incident photon flux augments, the corresponding increase in dark current will need to be addressed via the application of electron/hole blocking layers. The details of which materials provide the best performance has been extensively studied for x-ray based aSe devices \cite{Shiva1}, and will need to be explored in this application. No charge blocking layers were applied for the device under consideration.

For the boards used in this experiment, the characteristic spacing was confirmed by obtaining high resolution images using a Nikon Eclipse ME600 microscope paired with a Nikon DXM 1200 digital camera and image editing software {\tt Paint.NET}. The typical trace width and spaces were found to be 105.04 $\pm$ 1.94 $\mu$m and 146.57 $\pm$ 1.94 $\mu$m respectively.   These values are used when evaluating the applied electric field in the subsequent measurements described in Section \ref{sec:Results}. The trace heights are set by the manufacturing process and are 35 $\mu$m $\pm$ 5 $\mu$m.

Thermal evaporation deposition of selenium onto the boards was performed at Oak Ridge National Laboratory. An NRC/Varian 3117 E-Beam Vacuum Evaporator was retrofitted with a molybdenum boat to hold 722~mg of selenium pellets. The selenium pellets are purchased from Sigma-Aldrich \cite{pellets} and have a particle size $<$ 5 mm with a purity of selenium rated for $\geq$ 99.999 \%.  The PCBs were placed in a 3D printed mask 10 cm above the boat and the selenium was heated under high vacuum. The selenium coating was actively measured using a quartz monitor crystal with an Inficon XTM/2 deposition monitor to $\pm$1~nm precision to produce a 1.2~$\mu$m aSe layer.

%%%%%%%%%%%%%%%%%%%%%%%%%%%%%%%%%%%%%%%%%%%%%
%  The Cryogenic Temperature Test Stand here
%%%%%%%%%%%%%%%%%%%%%%%%%%%%%%%%%%%%%%%%%%%%%
\subsection{Cryogenic Temperature Test Stand}\label{sec:CryoStand}
\begin{figure}[htb]
    \centering
    \includegraphics[width=0.45\textwidth]{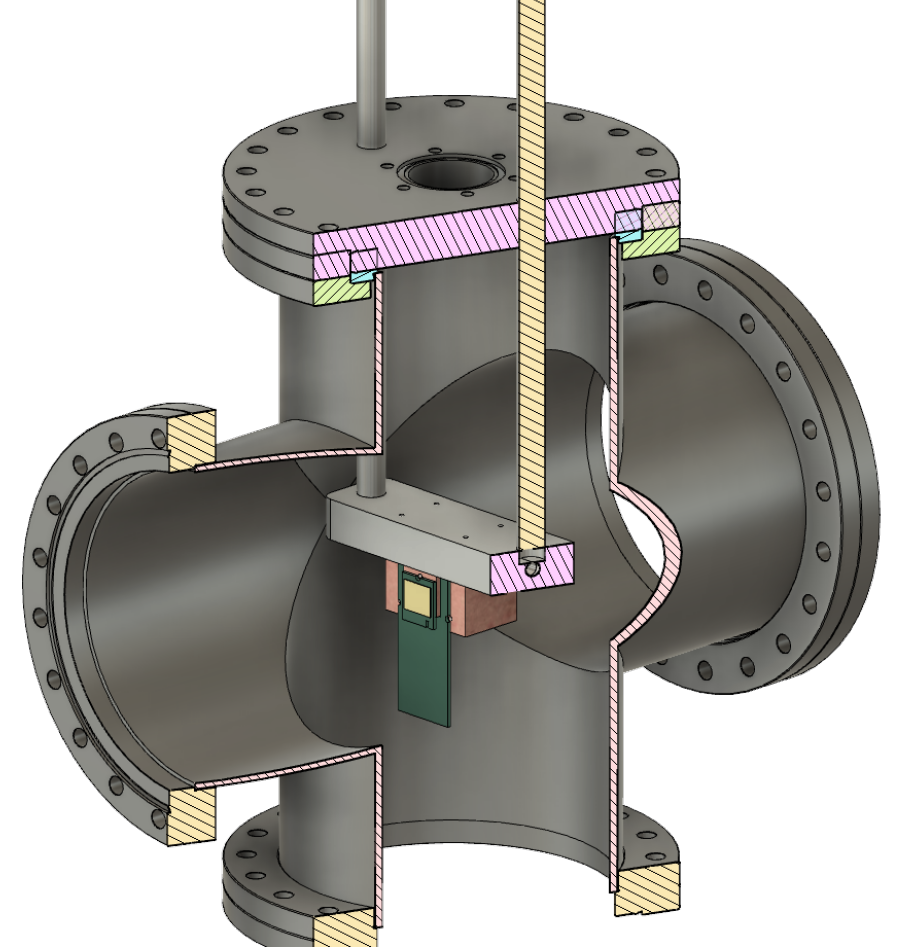}
    \includegraphics[scale=0.15]{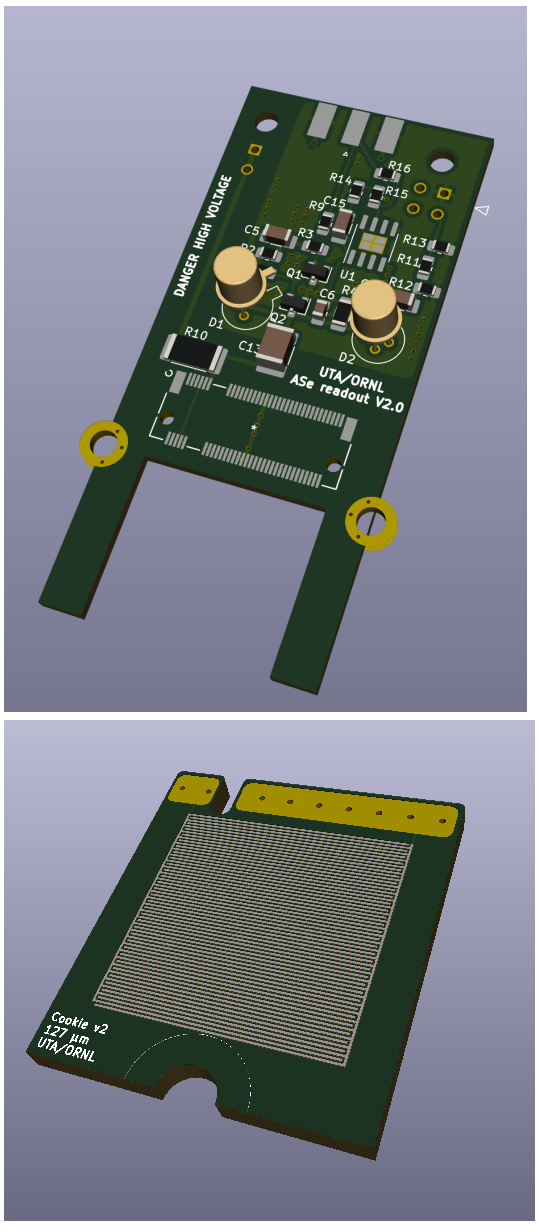}
    \caption{CAD model of the cryogenic teststand, custom carrier board, and test samples which were fabricated for use in characterization of the aSe devices.}
    \label{fig:cryostand}
\end{figure}

Testing the viability of the devices for noble elements detectors requires bringing the boards at liquid elements temperatures ($\sim 80$K). We achieve this task with the cryogenic test stand shown in Figure \ref{fig:cryostand}. The test stand is housed in a standard 8 in Conflat Flange (CF) cross (Lesker C-0800). The inner volume is evacuated via a turbo-molecular vacuum pump (Pfeiffer HiCube 80). A custom heat exchanger is fabricated from two 0.5~inch 304 stainless steel tubes with 0.125~inch walls which penetrate the top of the flange to allow the sample under test to be cooled. A block of 304 stainless steel allows the cryogenic fluid to circulate between the two tubes. The heat exchanger is cooled via a low pressure liquid nitrogen dewar, where the liquid is allowed to flow through the heat exchanger. In order to maintain flexibility with the setup, the sample holder is independent of the heat exchanger and mounts to the bottom of the heat exchanger. The sample holder used during the tests described here is machined from 101 copper and is bolted to the heat exchanger with a sheet of Indium (McMaster 8898N18) placed in between the heat exchanger and sample holder to aid in the thermal transfer. 

The test samples are mounted to the copper sample holder via a custom carrier PCB's which is outfitted with a standard M.2 connector, shown on the right of Figure \ref{fig:cryostand}. This connector is used for ease of testing various samples without interfering/moving the electronics. The sample PCB plugs into the M.2 connector on a carrier PCB which manages the connections for the readout electronics (shown schematically in Figure \ref{fig:cryostand} in green). 

\begin{figure}[htb]
    \centering
    \includegraphics[scale = 0.3]{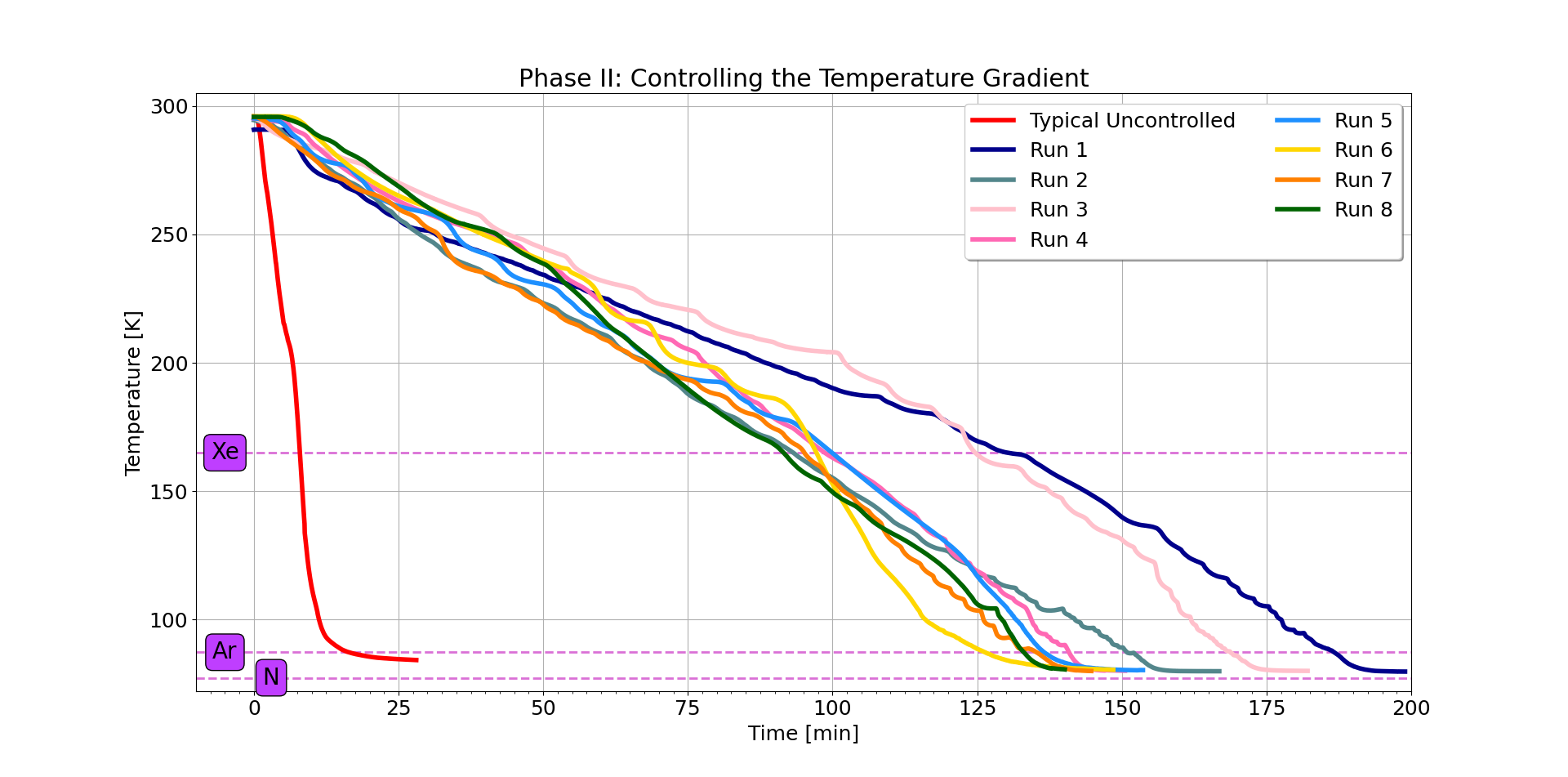}
    \caption{The data recorded from the PT-100 thin film resistive thermal devices during the cooldown of the experiment for the eight main data taking campaigns described below as well as one ``uncontrolled'' cooldown where the device was allowed to cool as fast as possible . The rate of temperature change was targeted to be between 1-2 Kelvin / minute during experimental operations. The relevant temperatures for various liquid cryogens (xenon, argon, and nitrogen) are noted on the plot for reference.}
    \label{fig:CooldownFigure}
\end{figure}

%https://www.ist-ag.com/sites/default/files/downloads/p0k1.232.6w.y.010_0.pdf
Two PT-100 thin film resistive thermal devices (RTD's) (P0K1.232.6W.Y.010 \cite{pt-100}) are mounted to monitor the temperature of the heat exchanger and the device under test. The rate of cooldown is determined by the flow of cyrogenic fluid through the heat exchanger. For the tests performed here, a manual valve was adjusted to maintain a cool down rate between 1.1 and 1.7 Kelvin/minute using liquid nitrogen. Figure \ref{fig:CooldownFigure} shows the typical cool-down curves over eight data runs compared to an uncontrolled cool-down where liquid nitrogen was allowed to flow at its maximum rate (resulting in achieving $< 80$K in under 30 mins). The samples can be kept at $\sim$80~K for extended periods of time by continuously flowing liquid nitrogen at a slow, but fixed rate.

The typical warm-up time varies slightly depending on the conditions in the lab, and is largely driven by the ambient temperature. Samples regularly reach room temperature within $\sim$10 hours after the nitrogen is shut off and the total data taking period per experiment lasting $\sim$40 hours in total.

%%%%%%%%%%%%%%%%%%%%%%%%%%%%%%%%%%%%%%%%%%%%%
%  The Data Acquisition and Readout Electronics here
%%%%%%%%%%%%%%%%%%%%%%%%%%%%%%%%%%%%%%%%%%%%%
\subsection{Data Acquisition and Readout Electronics}\label{sec:Electronics}

\begin{figure}[htb]
    \centering
    \includegraphics[width=0.6\textwidth]{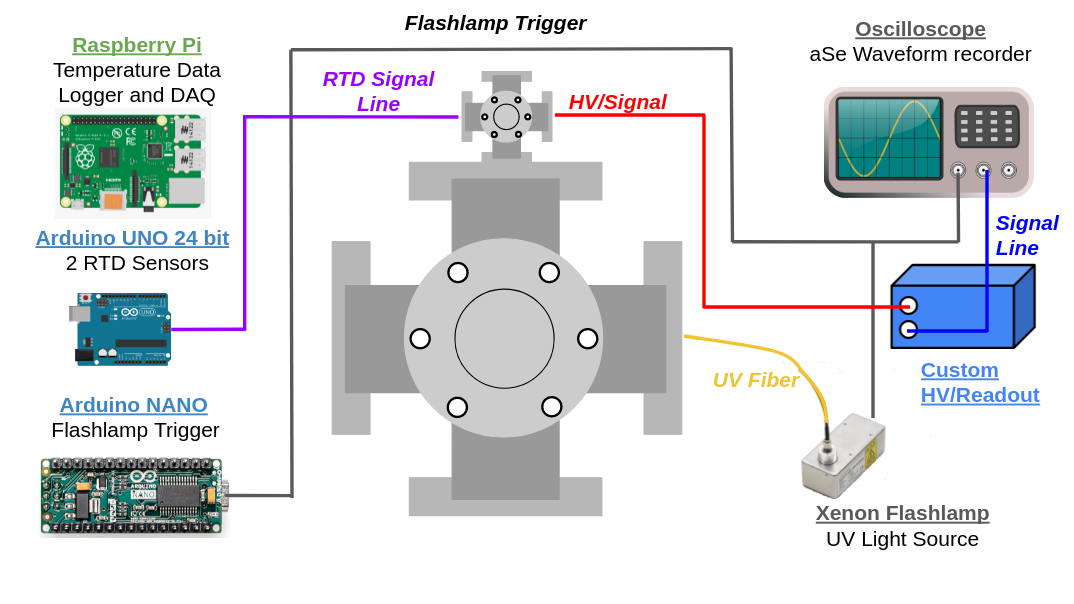}
    \includegraphics[width=0.25\textwidth]{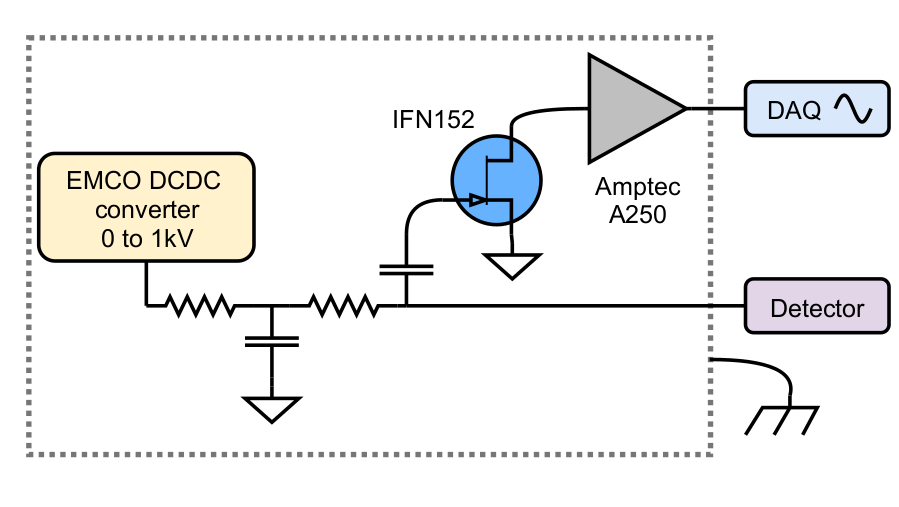}
    \caption{Schematic of the data acquisition system and readout electronics utilized in the characterization of the aSe device.}
    \label{fig:ReadoutSchematic}
\end{figure}

The data acquisition system and readout electronics are shown schematically in Figure \ref{fig:ReadoutSchematic}. The system is driven by a Raspberry Pi 3 Model B \cite{RPI} running a Python script which controls two Arduino's \cite{Arduino}. Two (P0K1.232.3K.B.010.M.U) RTD Platinum Sensors inside the cryogenic test stand are readout using the Arduino UNO coupled with a ARD-LTC2499 24-bit ADC data acquisition shield allowing the measured resistance to be converted to a temperature with milli-Kelvin precision.  The temperatures are recorded to an external solid state drive via a USB port on the Raspberry Pi. The RTD's have a dedicated Rigol DP832 power supply set at 2.048 V to match the ADC threshold. The Arduino Nano serves as a trigger for a 5 Watt Hamamatsu L11316-11 Xenon flashlamp \cite{Flashlamp} and LeCroy 6050 WaveRunner oscilloscope by providing a 5V signal with a rate configurable by the Raspberry Pi. The flashlamp has a dedicated PS-305D power supply set to 24 VDC. The oscilloscope triggers on the input from the flashlamp signal. Data files from the oscilloscope are stored on an external solid state drive. The temperature data and recorded waveforms are merged offline via the file number and the timestamp. 

In order to provide a high voltage (HV) bias between $\pm 750$ volts to the aSe board and readout the subsequent signals generated from its exposure to UV light, a custom HV/readout setup was implemented. The HV is generated via an ENCO DCDC converter which is powered by a 12V lead acid battery. The HV is filtered and applied to the sample. The charge is read out off the HV line via a decoupling capacitor and the signal is amplified with an Amptec A250 charge sensitive amplifier with an intrafet IFN152 as the input jfet \cite{amptek}. The output of the A250 is then sent to the oscilloscope for data collection. All electronics are housed in a shielded enclosure to further reduce noise. 

The high voltage supply was tested for stability over a 12 hour period and found to be stable to less than $0.1\%$ of the target value. The batteries on the readout box were regularly recharged to ensure no unexpected variation in the applied voltage. The xenon flashlamp output power was also tested for stability and repeatability by directly coupling the fiber optic to a THORLABS DET10A2 photodiode \cite{thorlabs} and found to have a `shot-to-shot' variation of $\sim 3\%$ (consistent with the lamps design document) and to have a consistent light output over a period of 12+ hours to within $1\%$.

%%%%%%%%%%%%%%%%%%%%%%%%%%%%%%%%%%%%%%%%%%%%%
%  The Results and Discussion here
%%%%%%%%%%%%%%%%%%%%%%%%%%%%%%%%%%%%%%%%%%%%%
\section{Results and Discussion}\label{sec:Results}

Table \ref{tab:RunningConditions} summarizes the nine data taking campaigns. The first campaign quantifies the effect of bulk trapping at various temperatures, referred to as ghosting. This phenomenon and the data used to understand its impact are described in Section \ref{sec:ghosting}. Six of the data campaigns were designed to test the response of the aSe device as a function of temperature at different applied electric fields. These results are discussed in Section \ref{sec:TempResults} and detailed numerical results are summarized in Appendix \ref{app:Temp}. Two data campaigns were taken to verify the repeatability of the results as a function of temperature and are described in the appendix Section \ref{app:repeat}. Variability in the results found during the repeat measurements is treated as a systematic on the results. The results in section \ref{sec:cryoTest} report the robustness tests against cryo-cycling. %Finally, these data are used to extract an ``effective'' electron and hole mobility as a function of temperature and field. The technique to extract the mobility and the results are discussed in Section \ref{sec:mobility}.

\begin{table}[htb]
    \centering
    \begin{tabular}{|c|c|c|c|}
    \hline 
         \textbf{Data Campaign} & \textbf{Applied Voltage} & \textbf{Electric Field} & \textbf{Charge Carrier}   \\
         & (Volts) & (Volts / $\mu$m) & \\
    \hline 
    \hline 
    Ghosting & \multirow{1}{*}{+400} & \multirow{1}{*}{+2.73} & \multirow{1}{*}{Electrons} \\
    \hline 
    \hline 
         & +400 &  +2.73 & Electrons \\
    \cline{2-4} 
         & -400 &  -2.73 & Holes \\
    \cline{2-4} 
    Temperature     & +530 &  +3.62 & Electrons \\
    \cline{2-4} 
    Characterization     & -530 &  -3.62 & Holes \\
    \cline{2-4}
         & +750 &  +5.16 & Electrons \\
    \cline{2-4} 
         & -750 &  -5.16 & Holes \\
    \hline 
    \hline 
         Repeatability & +400 &  +2.73 & Electrons \\
    \cline{2-4} 
         & -400 &  -2.73 & Holes \\
    \hline
    \end{tabular}
    \caption{Summary of the  data taking campaigns.}
    \label{tab:RunningConditions}
\end{table}

%%%%%%%%%%%%%%%%%%%%%%%%%%%%%%%%%%%%%%%%%%%%%
%  The Ghosting in aSe here
%%%%%%%%%%%%%%%%%%%%%%%%%%%%%%%%%%%%%%%%%%%%%
\subsection{Exposure Dependent Signal Reduction in aSe}\label{sec:ghosting}

The phenomenon of a change in the sensitivity of aSe based x-ray imaging detectors as a result of previous exposure to radiation is referred to as `ghosting' \cite{ghosting}. This phenomenon, which typically results in a decrease in sensitivity with subsequent exposures, has been determined to have the dominant mechanism due to bulk trapping of electrons which recombine subsequently with x-ray generated holes \cite{10.1117/12.465557}.  Holes may also become trapped in the aSe, affecting the response in either charge collection polarity. The typical lifetime, $\tau$, for a charge carrier to be released from a trap has the form \cite{doi:10.1063/1.368859} 
\begin{equation}\label{eqn:ghosting}
    \tau = \frac{\exp(E_{T}/kT)}{\nu}
\end{equation}
where $E_T$ is the energy depth of the trap (eestimated to be 0.9 eV above the valence band for holes and 1.2 eV below the conduction band edge for electrons \cite{doi:10.1080/09500838808214730}), $k$ is Boltzmann's constant, $T$ is the absolute temperature, and $\nu$ is the phonon frequency (taken as $10^{11}$~s$^{-1}$ \cite{10.1117/12.465557}). For room temperature operation $\tau$ has been found to be on the order of minutes for holes and hours for electrons. 

The overall impact that ghosting has on the performance of the aSe based detector is found to depend on both the applied electric field (reducing the effect of ghosting with increased field) and the time interval between exposures (with the effect of ghosting decreasing with longer time between exposures). This phenomenon has been observed in aSe detectors when exposed to x-rays and when the detector is in a vertical geometry \cite{Abbaszadeh2013}. 

\begin{figure}[htb]
    \centering
    \includegraphics[scale = 0.3]{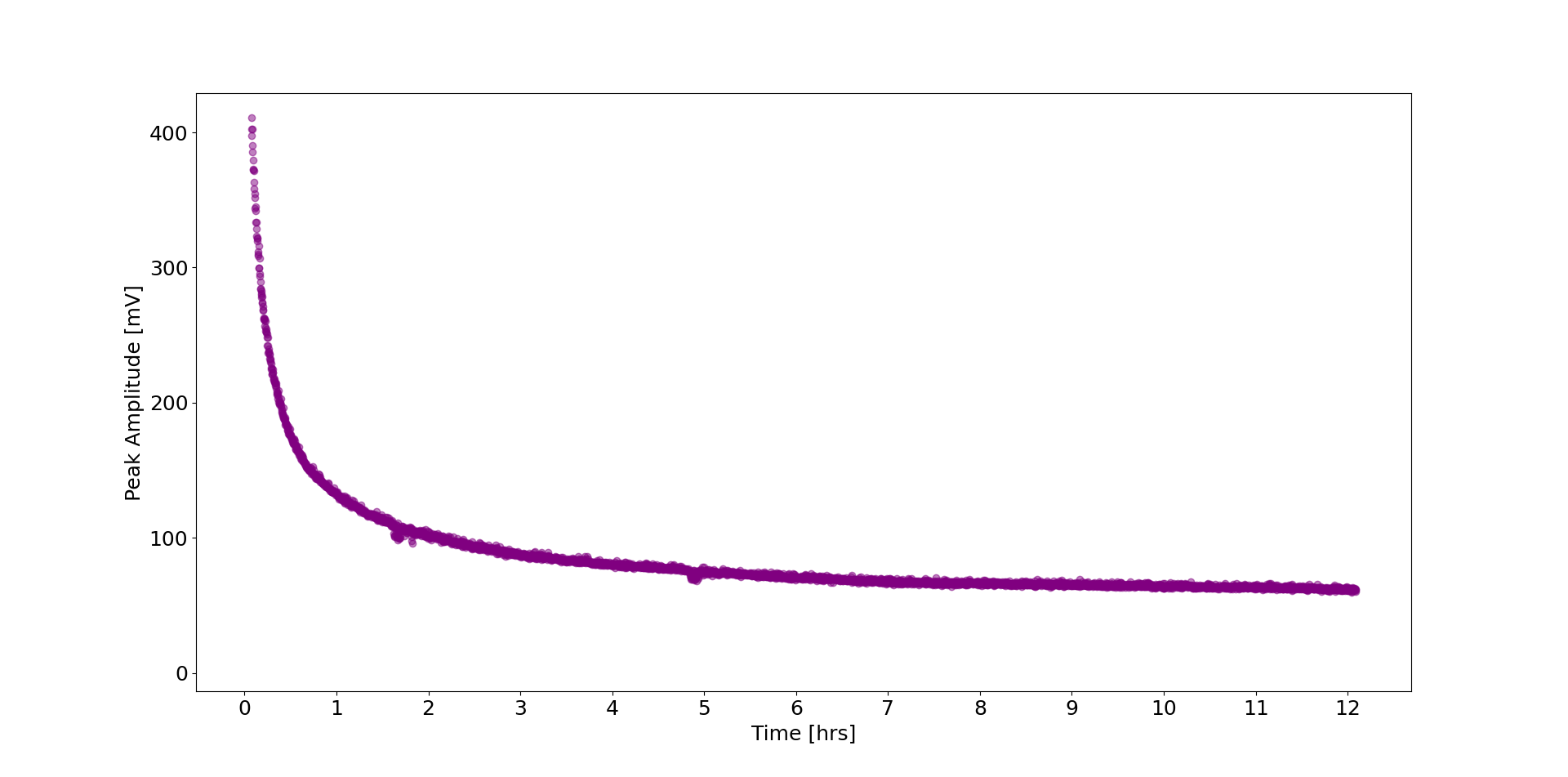}
    \includegraphics[scale = 0.3]{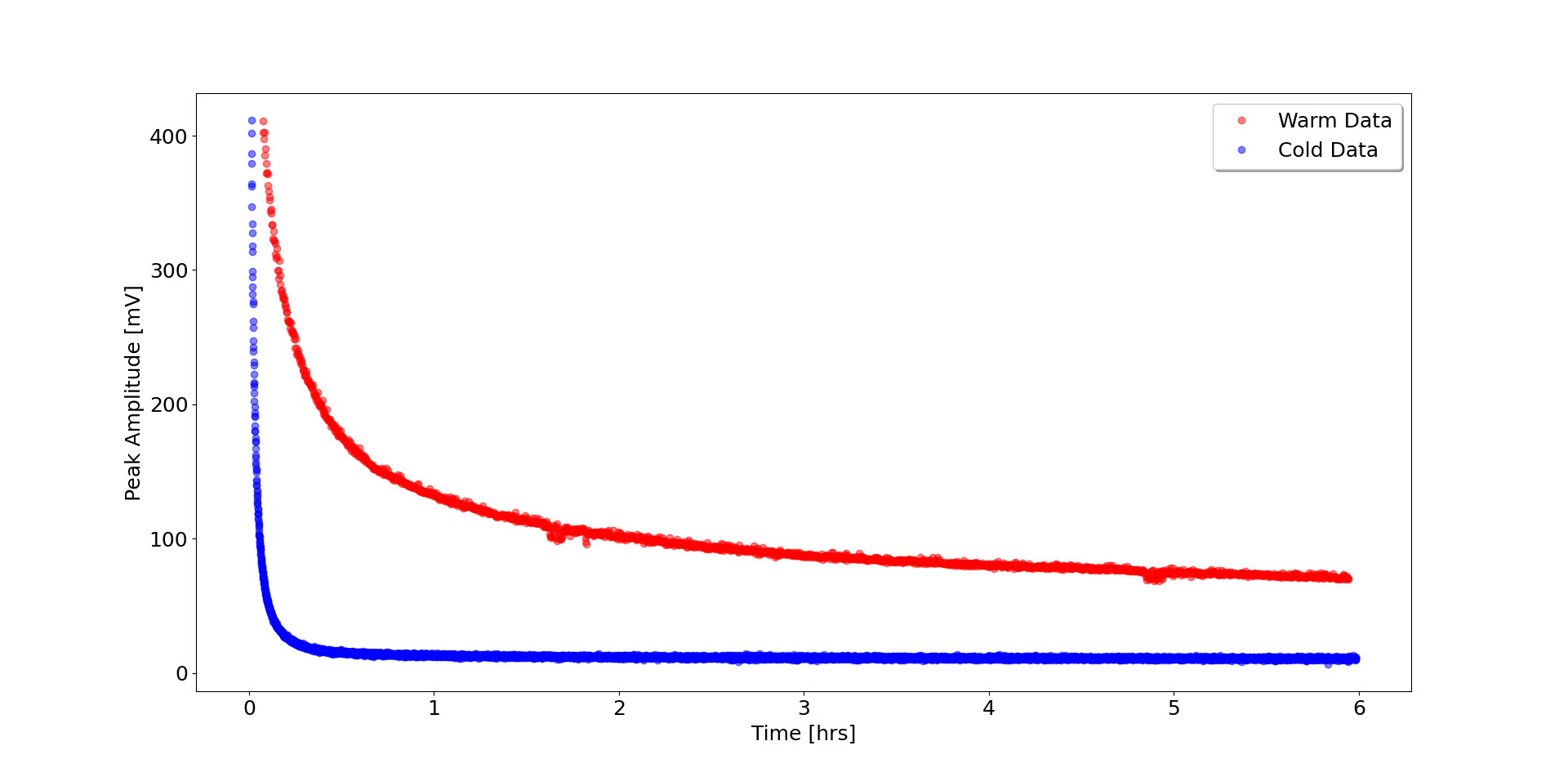}
    \caption{Top: The peak amplitude, defined in Section \ref{sec:DQandAnalysis}, at room temperature in the cryogenic temperature stand under vacuum and biased to +400 volts over a 12 hour data taking period when exposed to the xenon flashlamp every 10.8 seconds. The reponse of the aSe board can be seen to degrade over time until eventually reaching an equilibrium state after $\sim 6$ hours. Bottom: The recorded peak amplitude over a 6 hour period when the board is held at ambient temperature (red $\sim 290$~K) and when held at cryogenic temperature (blue $\sim 80$~K) in the cryogenic temperature stand under vacuum and biased to +400 volts and exposed to the xenon flash lamp every 10.8 seconds. The pulse amplitude drops much more quickly to the equilibrium state for the cryogenic temperature and can be interpreted as the longer lifetime for the charge traps associated with the ghosting effect.}
    \label{fig:Ghosting}
\end{figure}

We observe a reduction of the signal peak amplitude when exposing our windowless horizontal geometry detector to repeated pulses of light from the xenon lamp. This phenomenon suggests the effect is likely due to ghosting. The top of Figure \ref{fig:Ghosting} shows an example of the pulse amplitude recorded when the board is exposed to the Xenon flashlamp at a pulse rate of 0.1~Hz over a period of 12 hours. After a period of $\sim 6$ hours at room temperature, the system reaches an equilibrium state where the pulse amplitude is no longer noticeably changes. We attribute this to reaching a balance of the clearing of electron/hole traps at the given field and the creation of new traps with new electron/hole pairs created from the VUV light.

Given the application for the device under test, we explore this behavior in cold. Two dedicated runs were taken, the first at room temperature ($\sim 270$K) and the second at cryogenic temperature ($\sim 80$K). For both these runs the board was allowed to be at the designated temperature for a period of hours before being exposed to the Xenon flash lamp. The pulse amplitude, defined in Section \ref{sec:DQandAnalysis}, was then recorded over a six hour period. This data is shown in the bottom of Figure \ref{fig:Ghosting}. As anticipated, the time it takes to reach equilibrium at room temperature is longer than in cold. This is because the lifetime of the traps depends on the the temperature of the sample and becomes larger at lower temperatures, as shown in equation \ref{eqn:ghosting} The relative equilibrium pulse amplitude is different between the ``warm'' and ``cold'' data, but the stability of this equilibrium is similar. This is consistent with the model that thermal motion is what leads to the clearing of traps and thus at cryogenic temperatures becomes more pronounced.

\begin{figure}[htb]
    \centering
	\includegraphics[scale=0.3]{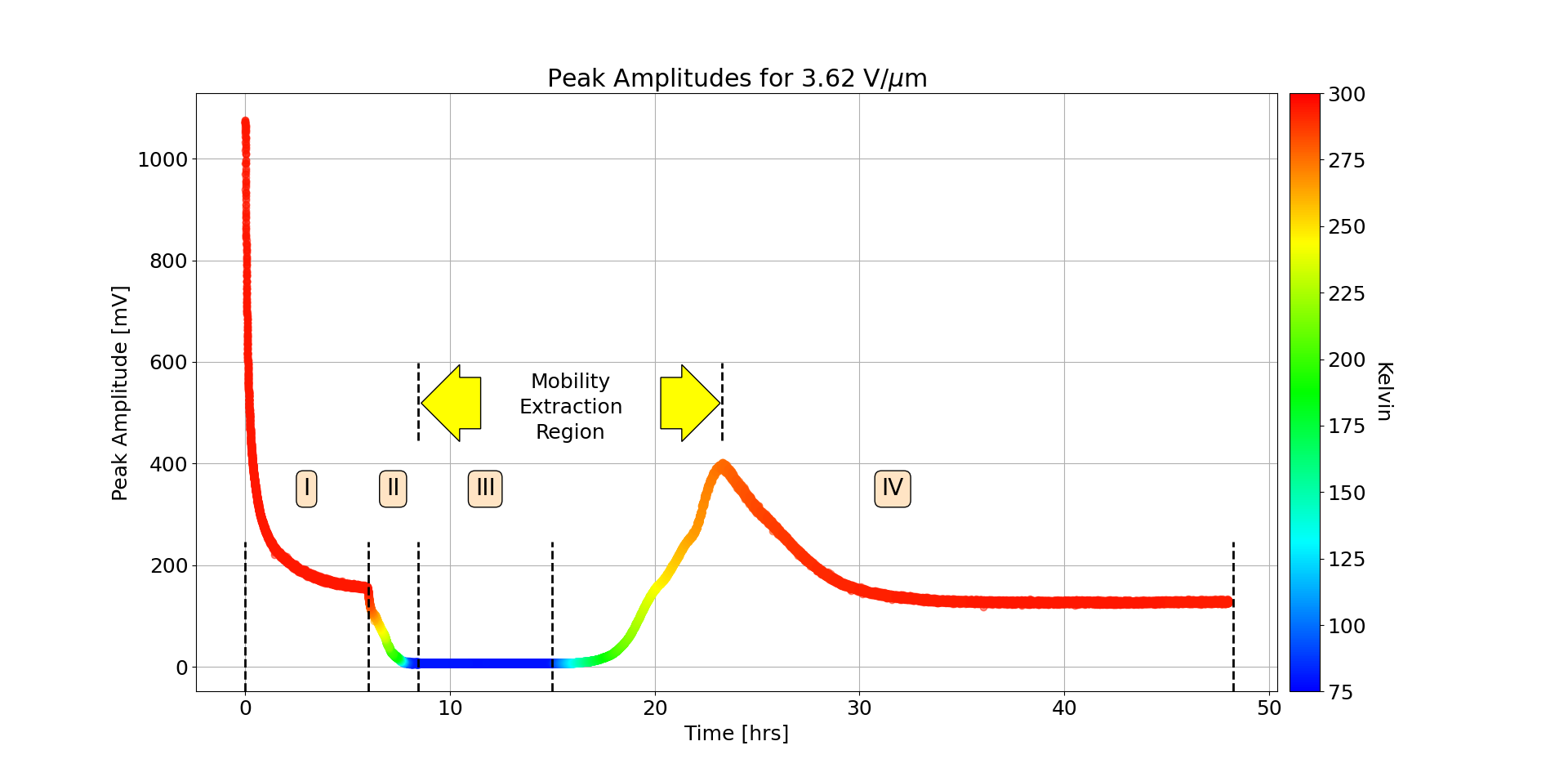} 
	\caption{Peak amplitudes with experimental phases marked for the mobility extraction region, ghosting mitigation (I), cooling down (II), lowest temperature (III) \& warming up (IV).}
    \label{fig:TypicalDataRun}
\end{figure}

To mitigate the impact of the ghosting effect across the measurements described below, we expose the system to UV light at a fixed frequency of 2~Hz for $\sim 6$ hours before beginning the cool down process. Moreover, we make the cool down process as slow as possible, typically ranging between two and three hours to allow for the system to stably transition. Finally, we remain at our lowest temperature we can achieve ($\sim 80$K) for a period of $\sim 7$ hours before allowing the system to warm up. We continue to take data until the system returns to the previous  peak amplitude equilibrium state when at room temperature before cooling down. The various stages described above are shown in Figure \ref{fig:TypicalDataRun}.

With the mitigation strategy in place to account for the effects of ghosting, Section \ref{sec:DQandAnalysis} describes the data quality cleanup and analysis procedures used.

%%%%%%%%%%%%%%%%%%%%%%%%%%%%%%%%%%%%%%%
%   Data Quality and Analysis Section
%%%%%%%%%%%%%%%%%%%%%%%%%%%%%%%%%%%%%%%
\subsection{Data Quality and Analysis Procedure}\label{sec:DQandAnalysis}

The data recorded using the setup described in Section \ref{sec:Electronics} has two files recorded. The first file is the waveform captured from the oscilloscope and the second is the temperature data recorded from the RTD's and saved on the Raspberry Pi. These two files are indexed such that they are matched in time and thus the data files are combined to provide a single file with both the recorded waveform and temperature.

\begin{figure}[htb]
    \centering
	\includegraphics[scale=0.2]{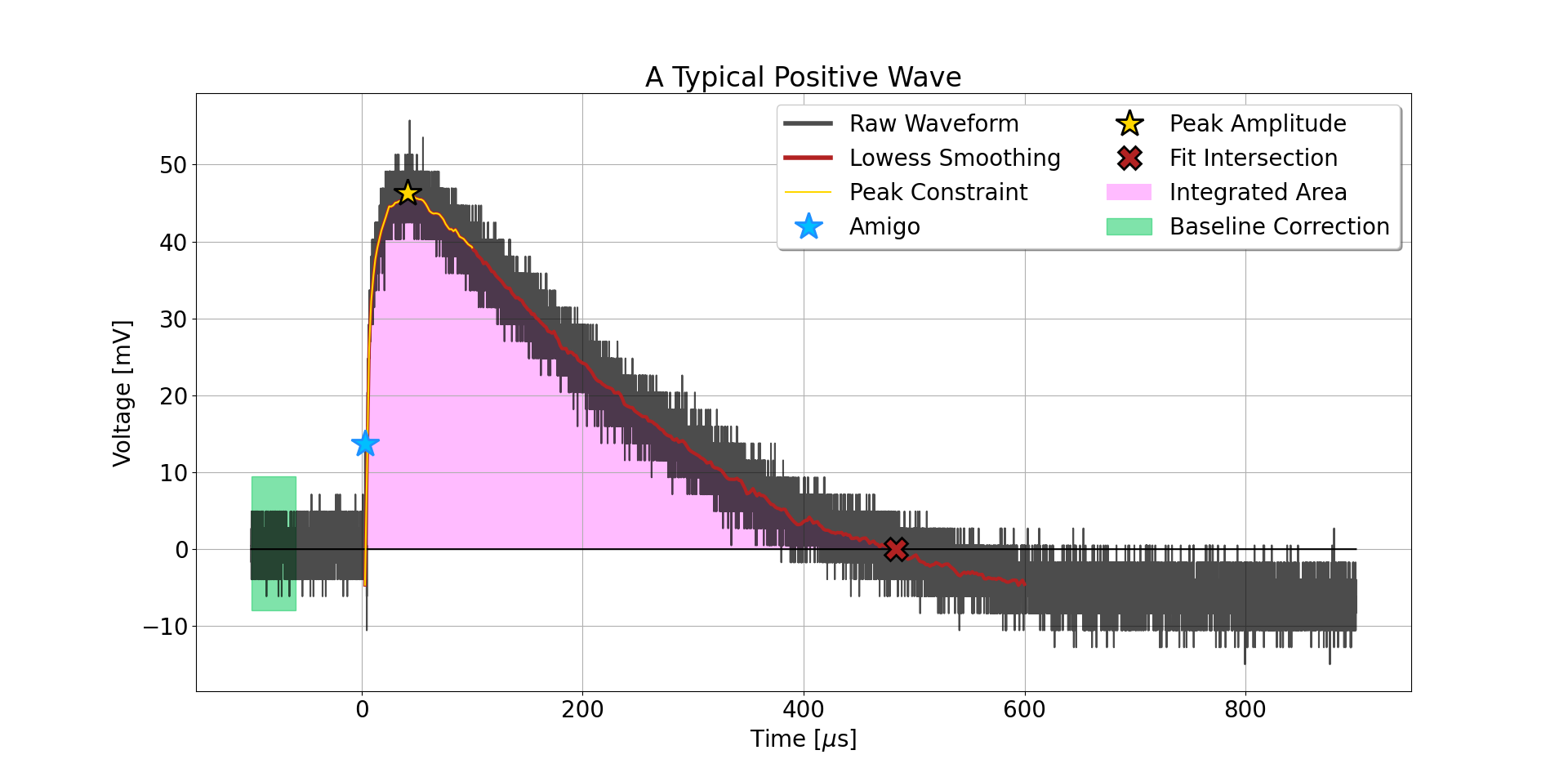} 
	\includegraphics[scale=0.2]{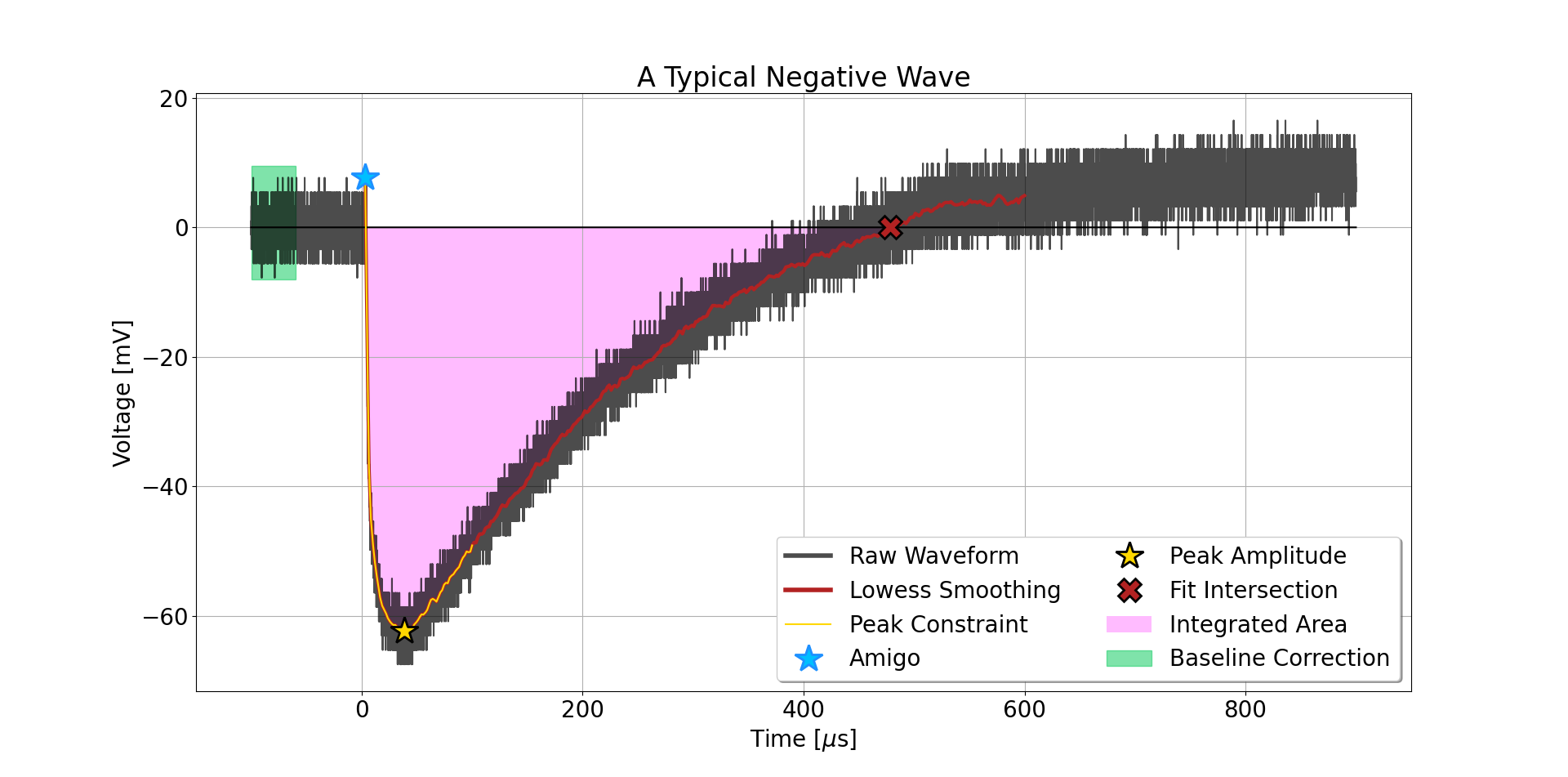}
	\includegraphics[scale=0.2]{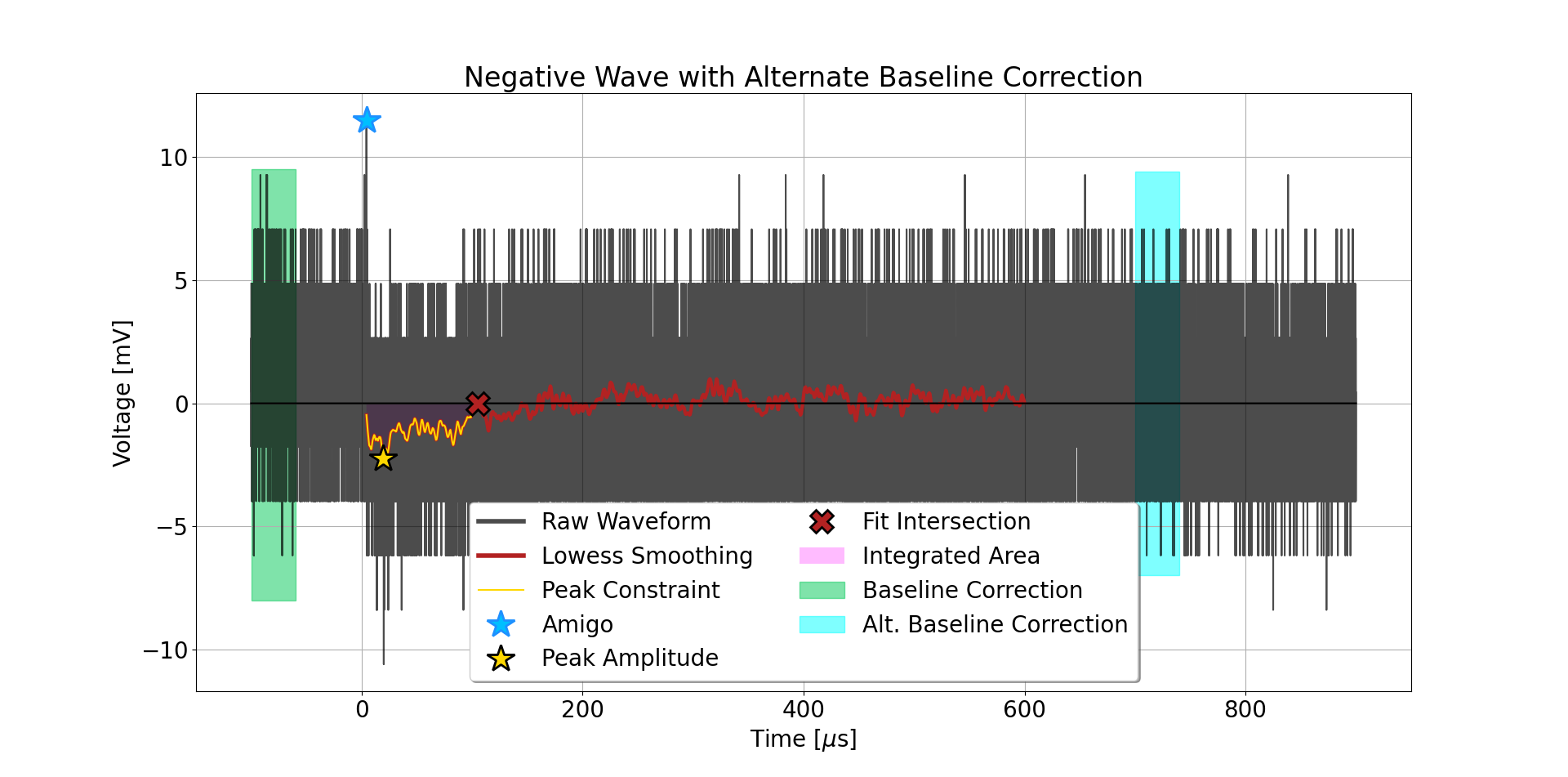}
	\caption{Example waveforms when collecting holes at ambient temperature (top), electrons at ambient temperature (middle), and electrons at cryogenic temperature (bottom). The figure highlights the methods described above of locating the start of the waveform via the characteristic inductive noise pickup (labeled ``Amigo'') from the start of the flashlamp, the effect of the smoothing algorithm are shown as the solid red line, the end of the pulse is identified as the ``intersection'', the peak amplitude, and integrated area are all illustrated. The bottom plot also illustrates the region of the waveform used when the alternative baseline correction is needed. }
    \label{fig:Illustration}
\end{figure}

Data quality was performed for all campaigns following the procedure described in this section. Applying positive polarity voltage resulted in ``positive waves'', while applying negative polarity voltage resulted in ``negative  waves''. Figure \ref{fig:Illustration} shows typical example waveforms in both cases. The illustrations are annotated to highlight the important features used to calculate the peak amplitude and area. 

To define the start time of each wave ($t_0$), we account for the known delay between the trigger pulse sent to the flashlamp and the actual formation of a light pulse. According to the flashlamp data sheet \cite{Flashlamp}, the delay relative to the input pulse is $\sim 4.0 - 4.5~\mu$s. The data show a characteristic ``pick-up'' due to inductive coupling between the flashlamp and the signal line for the aSe board. This peak (which is labeled `Amigo', as it provides a friendly reference point) appears reliably at $4.36~\mu$s after the trigger signal, thus defining our $t_0$.

Once the start of the waveform is defined, the waves are fitted using a LOcally WEighted Scatterplot Smoothing (LOWESS) \cite{moran_1984,statsmodels} statistical package. The fit ranges between  $t_0$ and $t = 600 ~\mu$s  providing a smoothed function of the waveform. The peak amplitude of the wave is found by sampling between $100 < t < 600~\mu$s and locating the minimum for negative waves or maximum for positive waves.  The area under the fit (integrated voltage) was calculated using the trapezoidal rule. Bounds of the integral were set by fixing the lower limit to 237 ns after the start of the fit and the upper limit to the fit and abscissa intersection following the peak amplitude. The lower limit was chosen to ensure integration occurs after any inductive noise seen in ``the Amigo'' has died out. The lower limit accounts for a documented delay and jitter time in the xenon flash lamp. 

Waveforms of particularly small negative amplitudes fail to return to baseline, and thus no area can be calculated. An alternate baseline correction is attempted in these cases by taking the baseline mean sufficiently far from where an intersection would occur ($700< t< 1100~\mu$s) and correcting that baseline. When the alternate baseline correction failed, the wave was removed from the data set. When the correction succeeded, the peak amplitude is recalculated, and the area found. Additionally, waves that produced integrated areas with an incorrect sign were removed from the data set. An illustration of the methods described above is given in Figure \ref{fig:Illustration}.

%Data analysis was completed using Python 3.7.11 running on an Intel Core i5-9400 CPU. Waves were shifted horizontally on the oscilloscope to begin at -100 ns. A baseline correction for a vertical offset of approximately -221.5 mV was made by taking the mean of the first 400 us of the wave and subtracting that value from the vertical data.

%For each data run, temperature and wave files were married and a list index was created for locating any wave in time and temperature. The index was masked below the start of phase III (9-10 hours) to omit waves subject to ghosting effects. A list index corresponding to a temperature bin of interest was generated and those waves were passed to an algorithm for removing bad wave forms. 

%The constancy of the electronic noise inherent in all waves was used to fix the starting point of the algorithm. A particular peak, dubbed Amigo, appears just before the rising edge, and is located by applying SciPy’s find peaks function between 4.01 us and 4.92 us []. The wave is then fit between Amigo and 600 us using statsmodels’ lowess smoothing function []. 

Values for peak amplitude and integrated areas are accumulated, averaged and the standard deviation of the set calculated. If an individual waveform is found to have a peak amplitude or integrated area greater than one standard deviation this waveform is removed as analysis shows these waveforms are typically saturated with external noise and thus shouldn't be considered. A new list index is then created keeping only waveforms which pass this filter. 

With the data cleanup completed, we extract the relevant physics from the remaining waveforms.

%%%%%%%%%%%%%%%%%%%%%%%%%%%%%%%%%%%%%%%%%%%%%
%  The Characterization across temperatures here
%%%%%%%%%%%%%%%%%%%%%%%%%%%%%%%%%%%%%%%%%%%%%
\subsection{Characterization across temperatures}\label{sec:TempResults}

To characterize the response of the aSe device, the peak amplitude (mV) and integrated area of the pulse (mV$\cdot \mu$s) are calculated for different temperature ranges. Within a given temperature range, 20 independent waveforms are averaged. The same procedure described above is used to calculate the peak amplitude, area and the standard deviation.

\begin{figure}[htb]
    \centering
    \includegraphics[scale=0.36]{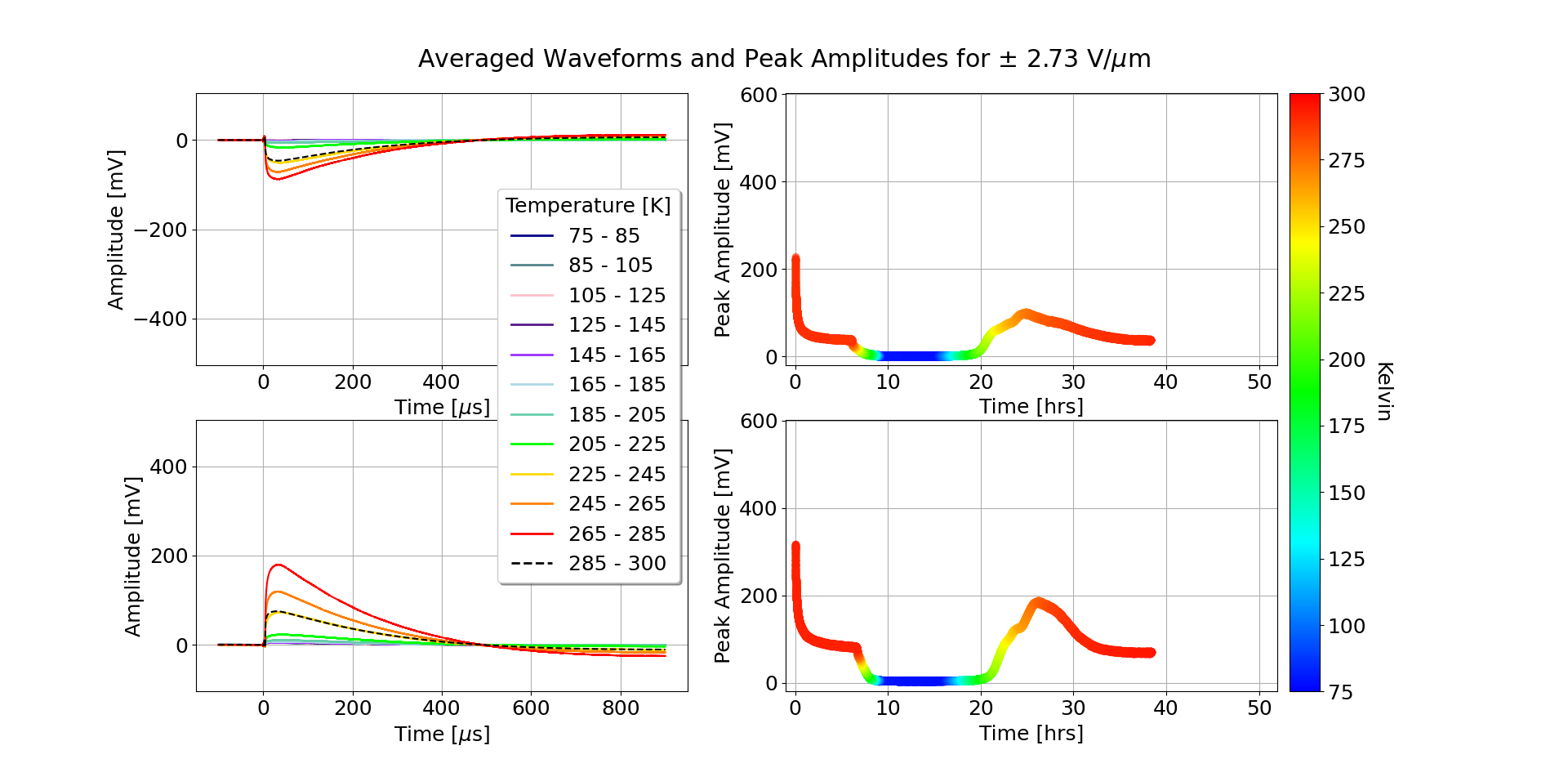}
    \caption{Left: The peak amplitude of the pulse recorded as over time during cooldown and warm up. Right: The average waveform as a function of temperature. The top row represent data for an applied voltage of -400 V (-2.73 V/$\mu$m electric field) and the bottom row represent data for an applied voltage of +400 V (+2.73 V/$\mu$ m electric field).}
    \label{fig:400Volts}
\end{figure}

\begin{figure}[htb]
    \centering
    \includegraphics[scale=0.36]{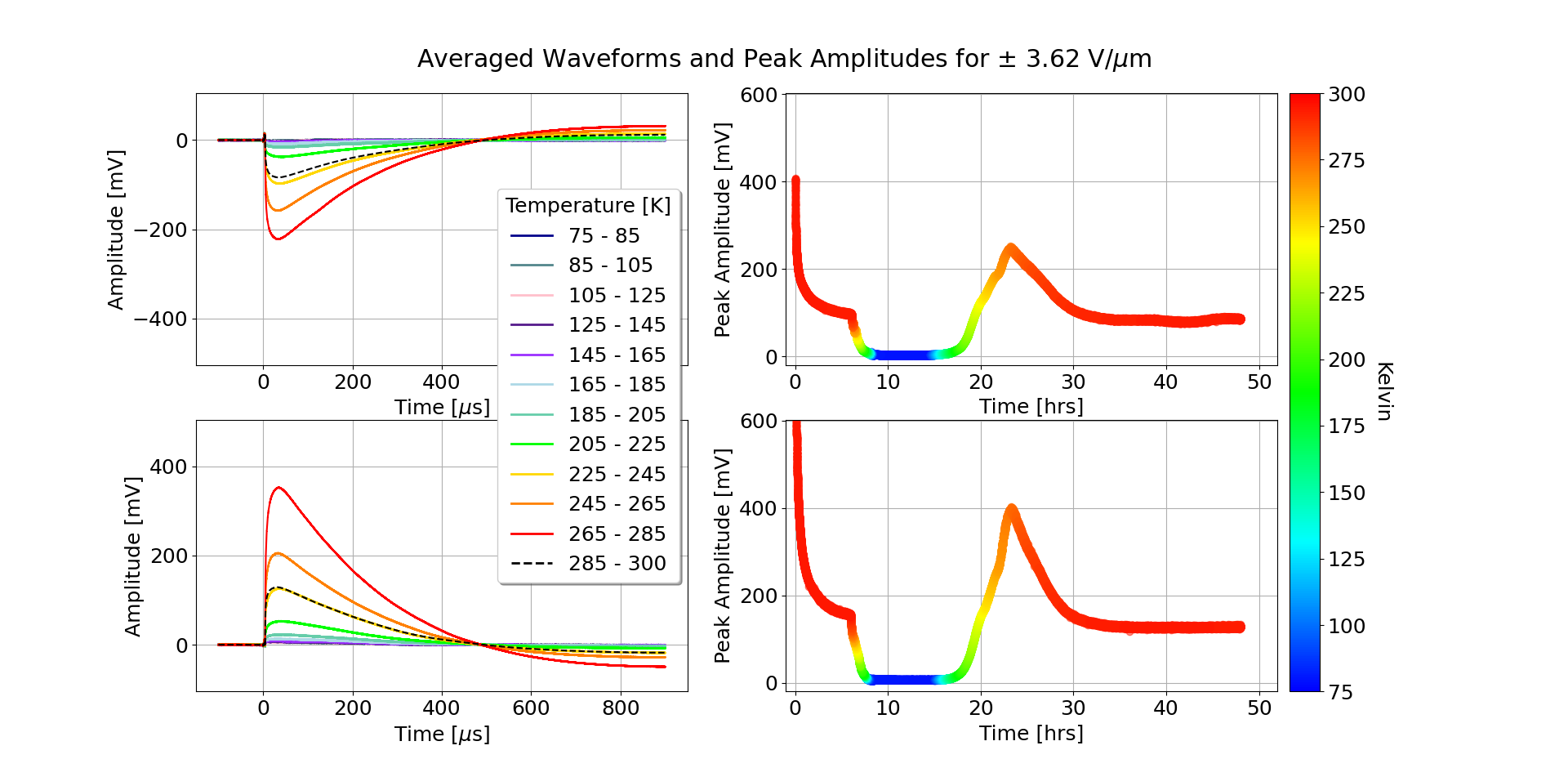}
    \caption{Left: The peak amplitude of the pulse recorded as over time during cooldown and warm up. Right: The average waveform as a function of temperature. The top row represent data for an applied voltage of -530 V (-3.62 V/$\mu$m electric field) and the bottom row represent data for an applied voltage of +530 V (+3.62 V/$\mu$ m electric field).}
    \label{fig:530Volts}
\end{figure}

\begin{figure}[htb]
    \centering
    \includegraphics[scale=0.36]{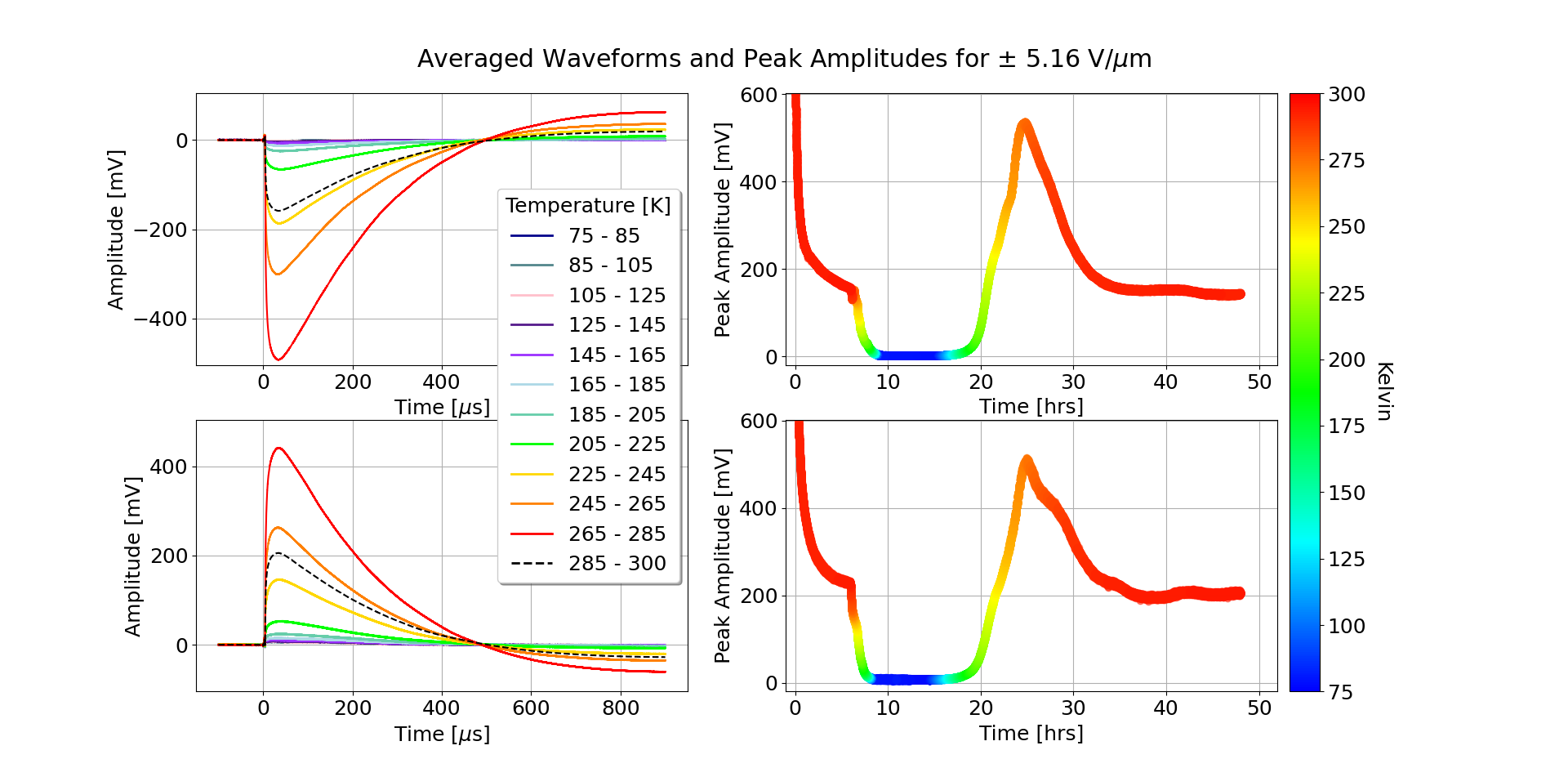}
    \caption{Left: The peak amplitude of the pulse recorded as over time during cooldown and warm up. Right: The average waveform as a function of temperature. The top row represent data for an applied voltage of -750 V (-5.16 V/$\mu$m electric field) and the bottom row represent data for an applied voltage of +750 V (+5.16 V/$\mu$ m electric field).}
    \label{fig:750Volts}
\end{figure}

Figures \ref{fig:400Volts}, \ref{fig:530Volts}, and \ref{fig:750Volts} show the results for three different applied fields on the peak amplitude as a function of temperature. These results are summarized in Tables \ref{tab:TempSummaryTable400V}, \ref{tab:TempSummaryTable530V}, \ref{tab:TempSummaryTable750V} in the appendix. A few general trends are observed from this data:
\begin{enumerate}
    \item While the magnitude of the peak amplitude is noticeably reduced at the lowest temperatures, it is definitively non-zero and has a pulse shape consistent with a response due to signal from the flashlamp
    
    \item The magnitude of the peak amplitude scales approximately with the size of the applied field, as would be expected. As an example ratio of the fields $\frac{3.62 \text{V}/\mu\text{m}}{2.73 \text{V}/\mu\text{m}} = 1.32$ and the ratio of the peak amplitudes at 265K-285K for those fields is 1.96 and between 75K-85K the ratio of the peak amplitudes is 1.42. Similar trends can be seen when looking at fields between $\frac{5.16 \text{V}/\mu\text{m}}{3.62 \text{V}/\mu\text{m}} = 1.43$ and the ratio of the peak amplitudes at 265K-285K for those fields is 1.42 and between 75K-85K the ratio of the peak amplitudes is 1.1. The data is summarized in Tables \ref{tab:TempSummaryTable400V}, \ref{tab:TempSummaryTable530V}, \ref{tab:TempSummaryTable750V}.
    
    \item The peak amplitude at the lowest temperatures is consistently higher when collecting electrons rather than holes. This trend holds true within the uncertainties of the measurement as the samples were warmed up. 
\end{enumerate}

It is worth noting that as the sample warms up, the effects due to ghosting dominate near 280 - 290~K. As the exposure to light in warm continues, the samples return to a similar equilibrium state to the start of the data taking, prior to cooling.

\clearpage

\begin{figure}[htb]
    \centering
    \includegraphics[scale = 0.19]{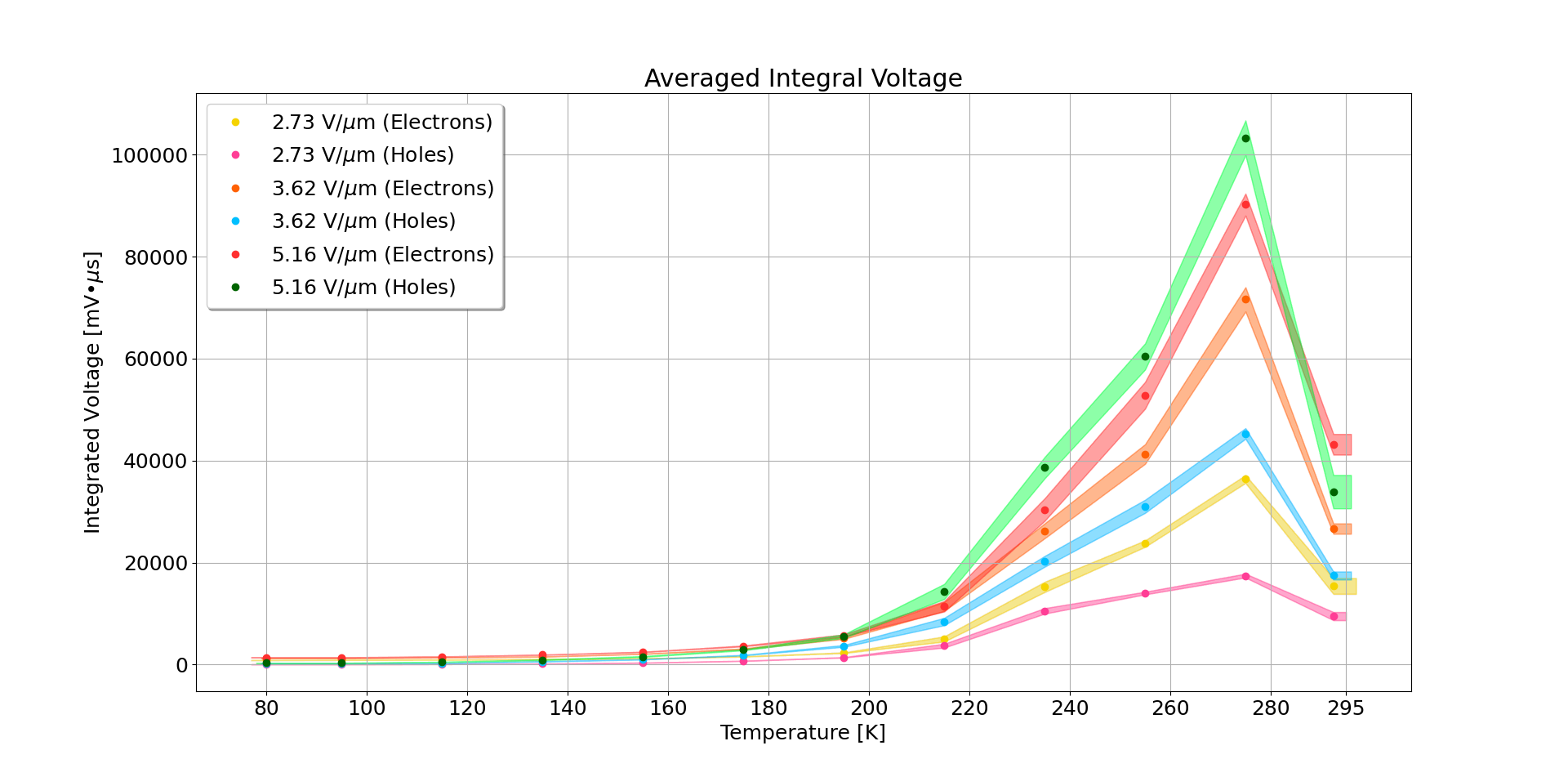}
    \includegraphics[scale = 0.19]{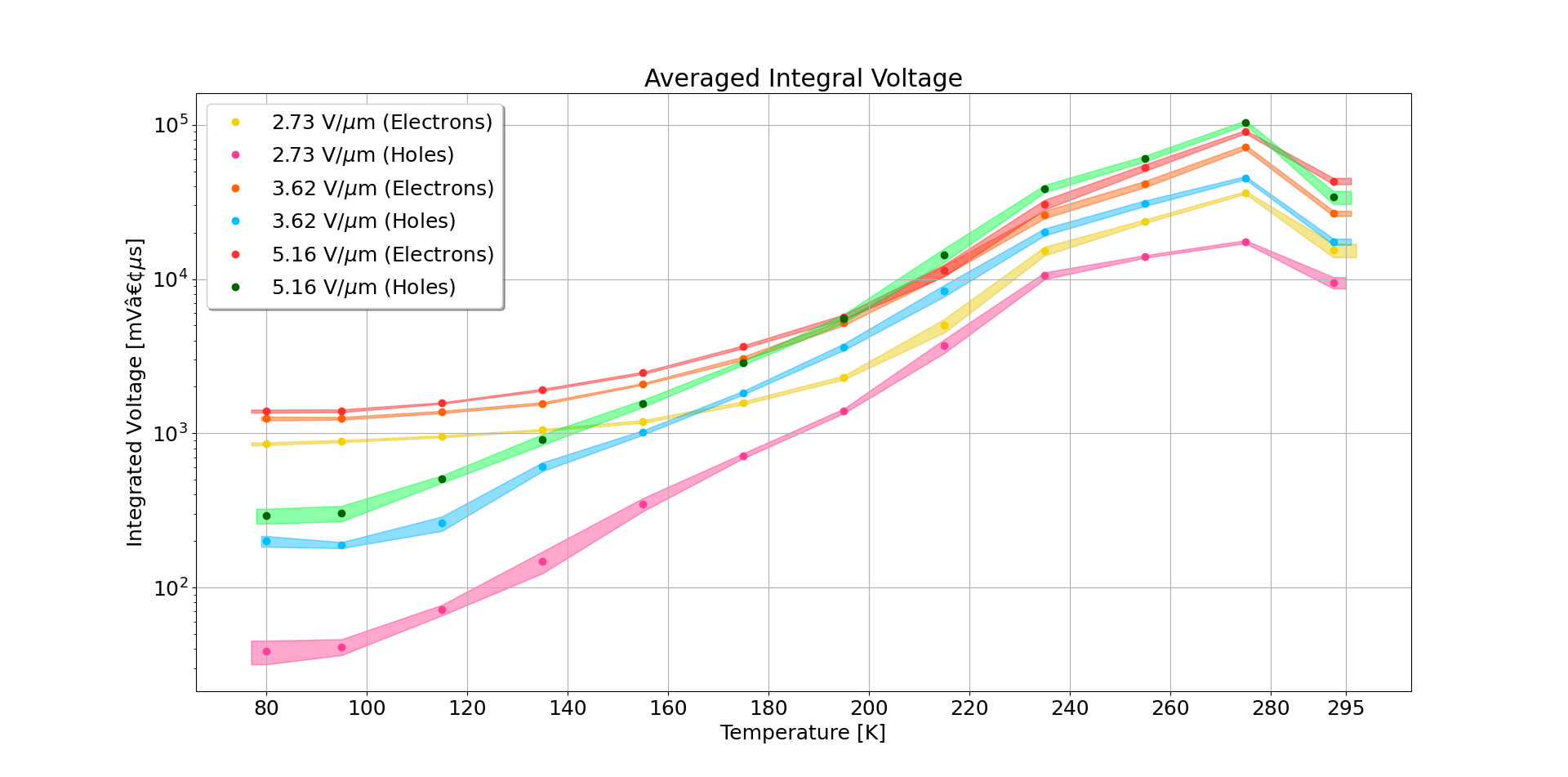}
    \caption{The pulse integral for as a function of temperature for all applied voltages plotted in linear (top) and log (bottom) scale. The bands represent the standard deviation of the measured integral when averaging together the waveforms within a given temperature bin.}
    \label{fig:IntegratedPulse}
\end{figure}

Figure \ref{fig:IntegratedPulse} shows the integrated pulse area as a function of temperature and applied voltage. These results are also summarized in Tables \ref{tab:TempSummaryTable400V}, \ref{tab:TempSummaryTable530V}, \ref{tab:TempSummaryTable750V} in the appendix. A similar set of observations can be seen in the amplitudes as was seen in the pulse areas. This consistency gives confidence that the same physics driving the pulse amplitude is present in the overall shape of the pulse, thus confirming that the signal is due to the response of the aSe detector.

\subsection{Usability in liquid noble element detectors}\label{sec:Usability}
This section describes the additional tests performed to ensure usability of amorphous selenium based devices in liquid noble elements time projection chambers. First, tests described in Section \ref{sec:cryoTest} demonstrate robustness of the prototypes against cryogenic cycling, checking if the deposited selenium remains on the board even after the exposure to extreme temperatures. Second, tests described in Section \ref{sec:purity} explore whether the introduction of aSe degrades argon purity.

%%%%%%%%%%%%%%%%%%%%%%%%%%%%%%%%%%%%%%%%%%%%%%%%
\subsubsection{Robustness against cryogenic cycling}\label{sec:cryoTest}
\begin{figure}[htb]
    \centering
    \includegraphics[width=0.75\textwidth]{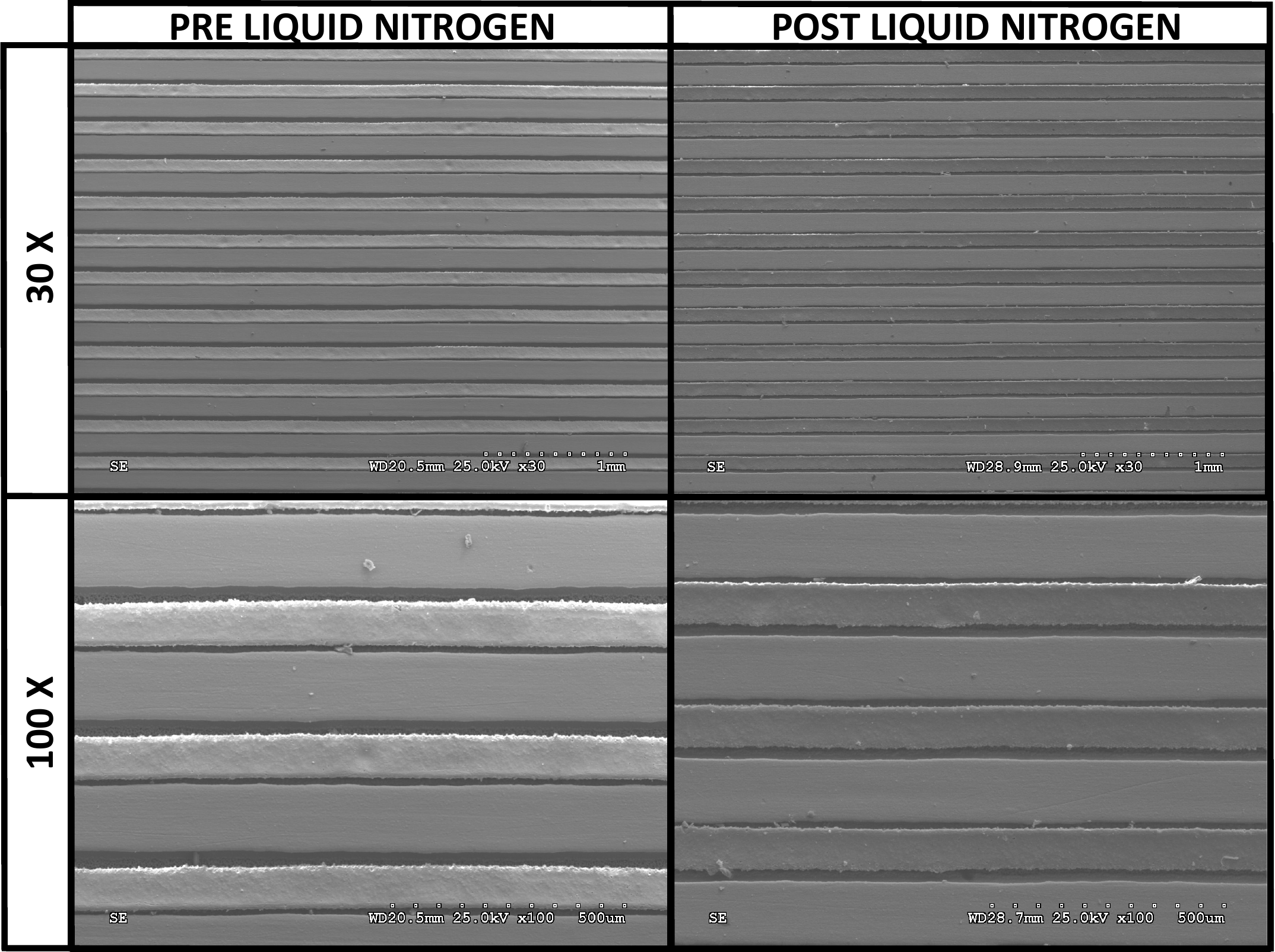}
    \caption{SEM images taken at 30x and 100x of the same region of an aSe board before and after cryocycling in LN$_{2}$. A scan of the board showed no noticeable defects of the aSe layer following cryocycling.}
    \label{fig:AFM}
\end{figure}

%%%%%%%%%%%%%%%%%%%%%%%%%%%%%%%%%%%%%%%%%%%%%%%
In addition to the repeated thermal cycling of these boards during the data taking campaigns -- after which no noticeable damage was observed -- the aSe coated boards were imaged using Scanning Electron Microscope (SEM) before and after submersion in a liquid nitrogen (LN$_{2}$) bath. The boards were lowered into the LN$_{2}$ bath from room temperature over a period of 10 mins with care taken to ensure no condensation formed on the board during the submersion. The LN$_{2}$ bath was then allowed to evaporate over a period of $>8$ hours and then imaged again afterwards. Examples of the before and after SEM images can be seen in Figure \ref{fig:AFM}. No apparent damage or cracking of the aSe layer can be seen in the SEM images and none could be seen during visual inspection. Taken together with the repeatable behavior of the aSe setup after multiple thermocycles provides confidence that such a windowless aSe detector is robust at the cryogenic temperatures of a liquid noble environment.

%%%%%%%%%%%%%%%%%%%%%%%%%%%%%%%%%%%%%%%%%%%%%%%%%
\subsubsection{Electronegative contaminants test}\label{sec:purity}
%%%%%%%%%%%%%%%%%%%%%%%%%%%%%%%%%%%%%%%%%%%%%%%%
A key feature of noble elements for particle detectors is the dual response to the passage of charged particles in the active volume in the form of correlated ionization charge and scintillation light. When developing new concepts for detectors intending to use both mechanisms, it is important to test that the light detection system does not suppress charge collection. 
The presence of electronegative contaminants in the liquid element, such as oxygen and water, is particularly pernicious since these molecules quench the charge produced by the ionizing radiation.   While noble element TPCs use hermetically sealed and leak-checked vessels to abate the leakage of external contaminants into the system, a sizable source of impurities can be introduced from the outgassing of internal surfaces.

We tested whether the outgassing of the aSe boards reaches levels harmful to charge collection by performing a measurement of the electron lifetime and water content at the Fermilab Material Test Stand (MTS)\cite{Andrews:2009zza} located at the Liquid Noble Test Facility (PAB).  The MTS is a 250 l liquid argon cryostat which allows 
to monitor the level of electronegative contaminants introduced by the material under examination  by positioning the material in the argon gas vapor (ullage) and submerging the material in the liquid. The MTS is equipped with an internal filtration system for oxygen and water contamination, which can be turn on or off as needed, and with a purity monitor to directly measure the effect of any material on the electron lifetime in the liquid argon.   

We examined a PCB board coated with 35 $\mu$m selenium on a surface area of 2$\times$2~cm$^2$. The board was cleaned by wiping the uncoated surface of the board with alcohol.  After the insertion of the board in the MTS, the sample chamber was purged with argon gas and evacuated several times to eliminate the contaminants acquired during insertion before the introduction in the active volume. 
Three runs were performed in the MTS. The first had no sample in the vessel and serves as a control to understand the behavior of the system when when the filters are on and off. The second has the sample suspended in the ullage where the effects of degradation to purity due to outgassing should be the most pronounced. Finally, the sample is lowered into liquid argon and left submerged. 

\begin{figure}[htb]
    \centering
    \includegraphics[width=0.75\textwidth]{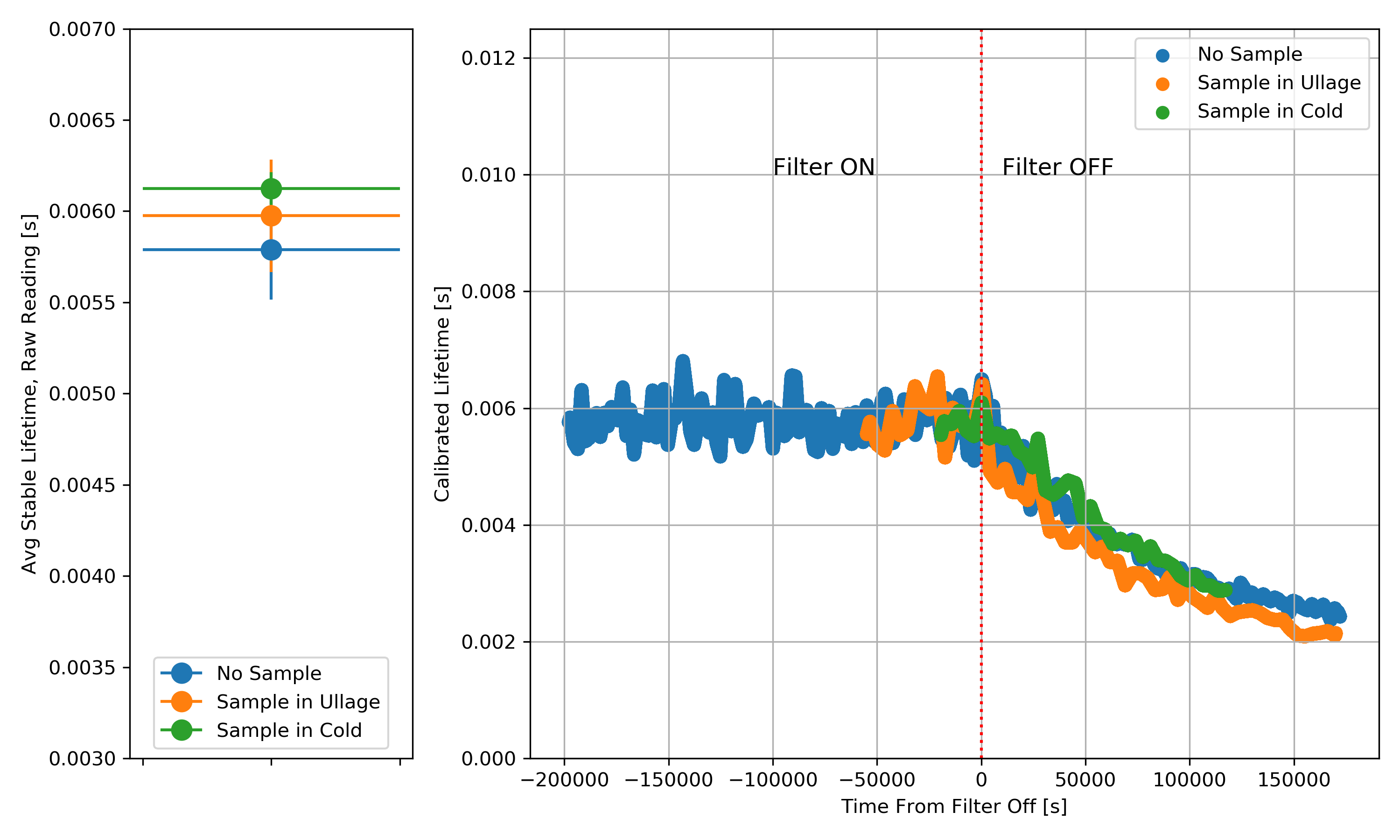}
    \caption{Left: average electron lifetime during active filtration period as read directly from the purity monitor (raw reading). The data is shown for the three run conditions: no sample (blue), sample in the ullage (orange), sample in the liquid (green). Right: calibrated lifetime as a function of time during the period when the filtration system is inactive. }
    \label{fig:eLifetime}
\end{figure}

To allow for comparison across runs, the same testing procedure is repeated for each run. First, the filtration system is activated, the electron lifetime is allowed to stabilize, and data is collected for several hours. Next, the filter system is switched off, and the decay of electron lifetime is observed. Figure \ref{fig:eLifetime} shows the results of the testing procedure for the three runs. The average lifetime during active filtration and the shape of the decay following the shut off of the filters are consistent across all runs. Thus, we conclude that the presence of the coated board does not suppress charge collection and does not negatively impact electron lifetime. %We also observed no significant water outgassing from the sample.

%%%%%%%%%%%%%%%%%%%%%%%%%%%%%%%%%%%%%%%%%%%%%
%  The Conclusions here
%%%%%%%%%%%%%%%%%%%%%%%%%%%%%%%%%%%%%%%%%%%%%
\section{Conclusions}\label{sec:conclusions}
In this paper, we have presented the response of a novel, windowless amorphous selenium based photon detector to UV light as a function of applied electric field and temperature. The device is constructed from low-cost commercially available printed circuit boards and simple thermal evaporation of selenium onto the board.

This initial exploration shows that such a device is: i) robust under cryogenic conditions, with the selenium remaining undamaged under cryogenic cycling and demonstrating the same performance after repeated thermal cycles, ii) responsive at cryogenic temperatures consistent with common liquid noble detectors (e.g. LAr and LXe),  iii) the response of the device is consistent with similar results shown for x-rays and gamma-rays (e.g. the observance of ghosting effects and the strength of the electron signal compared to the hole signal), and iv) preserving noble elements purity. 

Our finding that the device continues to respond at temperatures relevant to liquid noble detectors commonly used in high energy physics is particularly relevant to set future R\&D directions. While the flux of photons used in this experiment is quite high, we have provided a proof-of-principle demonstration that such a device could be be sensitive to a lower photon flux in a cryogenic environment, provided that a higher applied electric field is applied. Exploration into the response of such an aSe based device is ongoing with additional results expected to follow this work shortly.

\begin{figure}[htb]
    \centering
    \includegraphics[scale = 0.2]{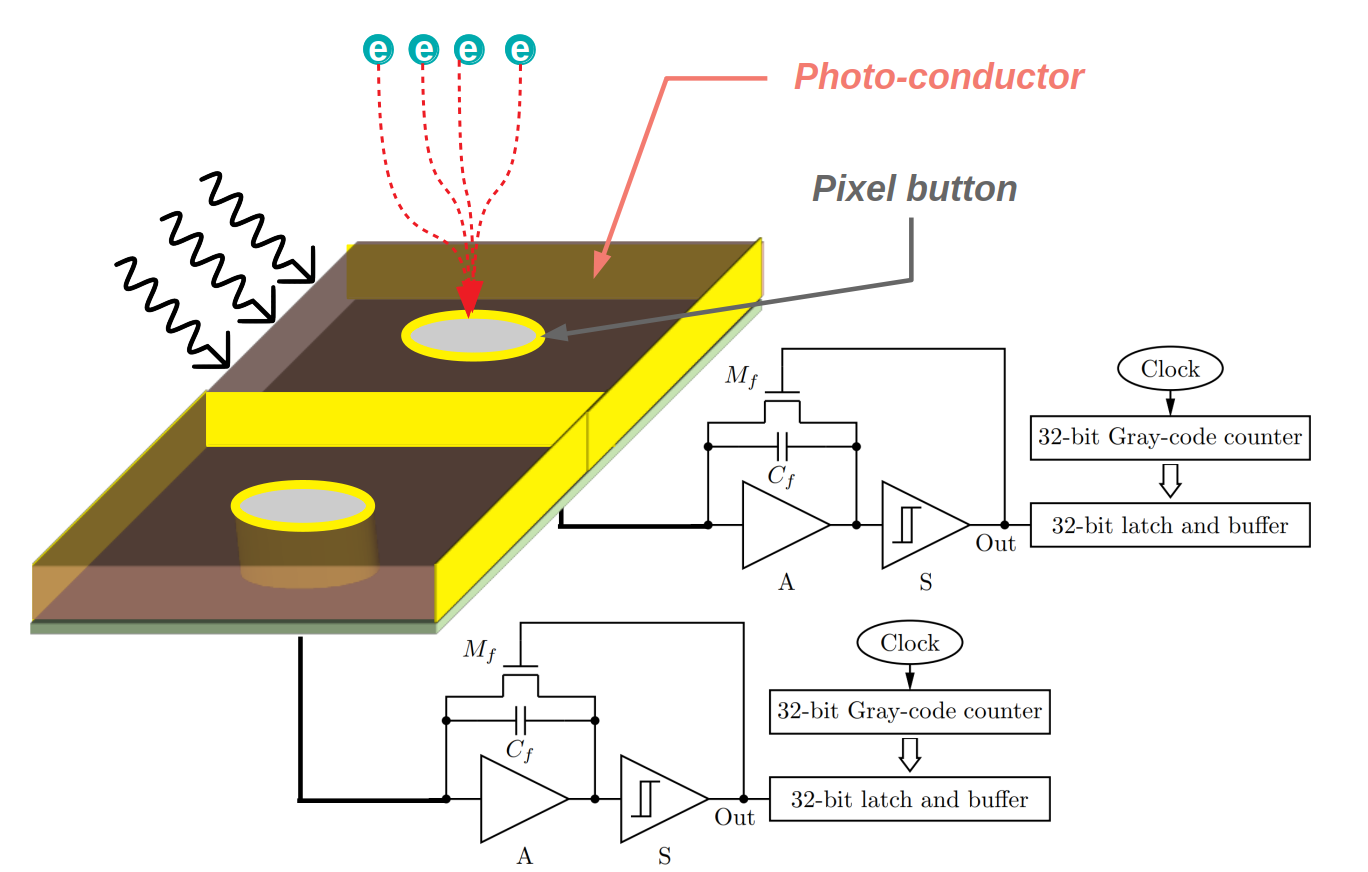}
    \caption{Conceptual sketch of an integrated charge and light (Q+L) sensor utilizing a windowless photoconductor, such as the device tested in this work, to directly detect the VUV photons produced in a noble element TPC. The conceptual device depicted here would use the same readout architecture used for the detection of ionization charge to detect the charge from the photoconductor. In this schematic, this is shown as the Q-Pix charge readout solution described in Reference \cite{1809.10213}.}
    \label{fig:QLSensor}
\end{figure}

A device based on the concept tested here opens the door to the possibility of making an integrated charge (Q) plus light (L) sensor, referred to as a ``Q+L sensor''. Such a sensor could simultaneously be sensitive to both the VUV photons produced in a liquid noble detector as well as the ionization charge created during the interaction of a charged particle with the noble element medium. A conceptual sketch of such a device using the Q-Pix \cite{1809.10213} charge readout architecture is shown in Figure \ref{fig:QLSensor}. Such a device, using amorphous selenium as the photoconductor, could have a large effective surface area, provide increased sensitivity to low energy physics and greater fidelity in energy reconstruction. 

The quantification of the improvement such a device will offer, as well as the realization of such a device in an experimental setup is the subject of future ongoing work.

%%%%%%%%%%%%%%%%%%%%%%%%%%%%%%%%%%%%%%%%%%%%%
%  The acknowledgments here
%%%%%%%%%%%%%%%%%%%%%%%%%%%%%%%%%%%%%%%%%%%%%
\acknowledgments
The authors would like to thank Scott Eichmann from the University of Texas Arlington machine shop for fabricating the cryogenic test stand. The authors also would like to thank Prof.~Shiva Abbaszadeh and Dr.~Katie Hellier from the University of California Santa Cruz for the aSe deposition of the boards used in the argon purity measurement. Thermal evaporation of the primary devices was done at Oak Ridge National Labs which is a DOE Office of Science User Facility. This material is based upon work supported by the U.S. Department of Energy, Office of Science, Office of High Energy Physics Award Number DE-SC0020065.
This document was prepared by the authors using the resources of the Fermi National Accelerator Laboratory (Fermilab), a U.S. Department of Energy, Office of Science, HEP User Facility. Fermilab is managed by Fermi Research Alliance, LLC (FRA), acting under Contract No. DE-AC02-07CH11359.

%%%%%%%%%%%%%%%%%%%%%%%%%%%%%%%%%%%%%%%%%%%%%
%  Push the appendix to a new page
%%%%%%%%%%%%%%%%%%%%%%%%%%%%%%%%%%%%%%%%%%%%%
\newpage
\appendix

%%%%%%%%%%%%%%%%%%%%%%%%%%%%%%%%%%%%
\section{Summary of pulse characterization across temperatures}\label{app:Temp}
%%%%%%%%%%%%%%%%%%%%%%%%%%%%%%%%%%%%%
Here we provide the corresponding data to the plots in Figures \ref{fig:400Volts}-\ref{fig:IntegratedPulse} associated with the pulse height and area as a function of temperature for the different applied fields. The errors quoted in these tables reflect the standard deviation from the averaging techniques described in Section \ref{sec:DQandAnalysis}.
%%%%%%%%%%%%%%%%%%%%%%%%%% 400 Volts %%%%%%%%%%%%%%%%%%%%%%%%%%%%%%%%%%%
\begin{table}[htb]
    \centering
    \begin{tabular}{|c|c|c|c|c|}
    \hline
        \textbf{Field} & \textbf{Charge Carrier} & \textbf{Temperature} & \textbf{Peak Amplitude} & \textbf{Integrated Pulse Area}  \\
        (V/$\mu$m) &  & (K) & (mV) & (mV$\cdot \mu$s) \\
    \hline
        \multirow{9}{*}{+2.73}& \multirow{9}{*}{Electrons} & 75-85 & 4.2 $\pm$ 0.1 & 854.7 $\pm$ 17.5 \\
              \cline{3-5}
              &           & 85-105 & 4.4 $\pm$ 0.0 & 890.1 $\pm$ 14.2 \\
              \cline{3-5}
              &           & 105-125 & 4.7 $\pm$ 0.1 & 953.7 $\pm$ 13.8 \\
              \cline{3-5}
              &           & 125-145 & 5.0 $\pm$ 0.0 & 1051.4 $\pm$ 12.4 \\
              \cline{3-5}
              &           & 145-165 & 5.4 $\pm$ 0.1 & 1191.9 $\pm$ 23.9 \\
              \cline{3-5}
              &           & 165-185 & 6.7 $\pm$ 0.2 & 1574.2 $\pm$ 34.5 \\
              \cline{3-5}
              &           & 185-205 & 10.1 $\pm$ 0.5 & 2296.7 $\pm$ 66.2 \\
              \cline{3-5}
              &           & 205-225 & 22.5 $\pm$ 2.4 & 4996.6 $\pm$ 499.2 \\
              \cline{3-5}
              &           & 225-245 & 72.4 $\pm$ 5.3 & 15204.2 $\pm$ 994.6 \\
              \cline{3-5}
              &           & 245-265 & 119.2 $\pm$ 3.4 & 23695.2 $\pm$ 622.6 \\
              \cline{3-5}
              &           & 265-285 & 179.4 $\pm$ 4.0 & 36369.9 $\pm$ 762.2 \\
              \cline{3-5}
              &           & 285-300 & 75.2 $\pm$ 7.1 & 15389.1 $\pm$ 1547.9 \\
    \hline
        \multirow{9}{*}{-2.73}& \multirow{9}{*}{Holes} & 75-85 & 1.3 $\pm$ 0.2 & 38.5 $\pm$ 6.7 \\
              \cline{3-5}
              &           & 85-105 & 1.3 $\pm$ 0.1 & 41.3 $\pm$ 4.9 \\
              \cline{3-5}
              &           & 105-125 & 1.6 $\pm$ 0.1 & 71.6 $\pm$ 5.6 \\
              \cline{3-5}
              &           & 125-145 & 1.9 $\pm$ 0.1 & 147.4 $\pm$ 23.8 \\
              \cline{3-5}
              &           & 145-165 & 2.4 $\pm$ 0.1 & 345.7 $\pm$ 33.7 \\
              \cline{3-5}
              &           & 165-185 & 3.4 $\pm$ 0.1 & 715.1 $\pm$ 26.1 \\
              \cline{3-5}
              &           & 185-205 & 6.1 $\pm$ 0.3 & 1394.3 $\pm$ 49.2 \\
              \cline{3-5}
              &           & 205-225 & 16.8 $\pm$ 1.8 & 3682.3 $\pm$ 373.8 \\
              \cline{3-5}
              &           & 225-245 & 51.0 $\pm$ 2.9 & 10505.1 $\pm$ 541.2 \\
              \cline{3-5}
              &           & 245-265 & 71.6 $\pm$ 1.8 & 14000.7 $\pm$ 263.8 \\
              \cline{3-5}
              &           & 265-285 & 87.7 $\pm$ 2.2 & 17431.7 $\pm$ 380.7 \\
              \cline{3-5}
              &           & 285-300 & 46.4 $\pm$ 4.0 & 9486.1 $\pm$ 804.7 \\
    \hline
    \end{tabular}
    \caption{Summary of the mean peak amplitude and integrated pulse area across the temperature range probed for an electric field of $\pm$ 2.73 V/$\mu$m.}
    \label{tab:TempSummaryTable400V}
\end{table}

%%%%%%%%%%%%%%%%%%%%%%%%%% 530 Volts %%%%%%%%%%%%%%%%%%%%%%%%%%%%%%%%%%%
\begin{table}[htb]
    \centering
    \resizebox{0.98\textwidth}{!}{%
    \begin{tabular}{|c|c|c|c|c|}
    \hline
        \textbf{Field} & \textbf{Charge Carrier} & \textbf{Temperature} & \textbf{Peak Amplitude} & \textbf{Integrated Pulse Area}  \\
        (V/$\mu$m) &  & (K) & (mV) & (mV$\cdot \mu$s) \\
    \hline
        \multirow{9}{*}{+3.62}& \multirow{9}{*}{Electrons} & 75-85 & 6.0 $\pm$ 0.1 & 1247.3 $\pm$ 33.7 \\
              \cline{3-5}
              &           & 85-105 & 6.0 $\pm$ 0.1 & 1249.5 $\pm$ 24.5 \\
              \cline{3-5}
              &           & 105-125 & 6.4 $\pm$ 0.1 & 1373.8 $\pm$ 23.4 \\
              \cline{3-5}
              &           & 125-145 & 7.2 $\pm$ 0.1 & 1562.6 $\pm$ 25.9 \\
              \cline{3-5}
              &           & 145-165 & 9.0 $\pm$ 0.2 & 2087.5 $\pm$ 25.8 \\
              \cline{3-5}
              &           & 165-185 & 12.9 $\pm$ 0.5 & 3057.1 $\pm$ 87.1 \\
              \cline{3-5}
              &           & 185-205 & 22.8 $\pm$ 1.1 & 5237.1 $\pm$ 224.5 \\
              \cline{3-5}
              &           & 205-225 & 52.4 $\pm$ 4.3 & 11381.6 $\pm$ 881.1 \\
              \cline{3-5}
              &           & 225-245 & 126.4 $\pm$ 7.8 & 26203.4 $\pm$ 1456.2 \\
              \cline{3-5}
              &           & 245-265 & 205.2 $\pm$ 9.6 & 41276.3 $\pm$ 1923.2 \\
              \cline{3-5}
              &           & 265-285 & 351.9 $\pm$ 12.0 & 71610.6 $\pm$ 2387.3 \\
              \cline{3-5}
              &           & 285-300 & 129.1 $\pm$ 4.4 & 26652.3 $\pm$ 1013.4 \\
    \hline
        \multirow{9}{*}{-3.62}& \multirow{9}{*}{Holes} & 75-85 & 2.8 $\pm$ 0.1 & 199.7 $\pm$ 15.9 \\
              \cline{3-5}
              &           & 85-105 & 2.7 $\pm$ 0.1 & 188.4 $\pm$ 8.3 \\
              \cline{3-5}
              &           & 105-125 & 3.3 $\pm$ 0.1 & 260.9 $\pm$ 27.8 \\
              \cline{3-5}
              &           & 125-145 & 4.3 $\pm$ 0.1 & 607.7 $\pm$ 39.5 \\
              \cline{3-5}
              &           & 145-165 & 5.7 $\pm$ 0.1 & 1010.8 $\pm$ 33.6 \\
              \cline{3-5}
              &           & 165-185 & 9.0 $\pm$ 0.3 & 1832.7 $\pm$ 49.3 \\
              \cline{3-5}
              &           & 185-205 & 16.2 $\pm$ 0.9 & 3631.7 $\pm$ 174.2 \\
              \cline{3-5}
              &           & 205-225 & 37.7 $\pm$ 3.3 & 8415.9 $\pm$ 721.7 \\
              \cline{3-5}
              &           & 225-245 & 97.7 $\pm$ 5.9 & 20223.3 $\pm$ 1055.2 \\
              \cline{3-5}
              &           & 245-265 & 158.0 $\pm$ 6.2 & 31014.0 $\pm$ 1284.7 \\
              \cline{3-5}
              &           & 265-285 & 222.4 $\pm$ 5.6 & 45310.1 $\pm$ 1037.2 \\
              \cline{3-5}
              &           & 285-300 & 84.0 $\pm$ 4.0 & 17456.3 $\pm$ 787.4 \\
    \hline
    \end{tabular}}
    \caption{Summary of the mean peak amplitude and integrated pulse area across the temperature range probed for an electric field of $\pm$ 3.62 V/$\mu$m.}
    \label{tab:TempSummaryTable530V}
\end{table}

%%%%%%%%%%%%%%%%%%%%%%%%%% 750 Volts %%%%%%%%%%%%%%%%%%%%%%%%%%%%%%%%%%%
\begin{table}[htb]
    \centering
    \resizebox{0.98\textwidth}{!}{%
    \begin{tabular}{|c|c|c|c|c|}
    \hline
        \textbf{Field} & \textbf{Charge Carrier} & \textbf{Temperature} & \textbf{Peak Amplitude} & \textbf{Integrated Pulse Area}  \\
        (V/$\mu$m) &  & (K) & (mV) & (mV$\cdot \mu$s) \\
    \hline
        \multirow{9}{*}{+5.16}& \multirow{9}{*}{Electrons} & 75-85 & 6.6 $\pm$ 0.1 & 1390.9 $\pm$ 31.8 \\
              \cline{3-5}
              &           & 85-105 & 6.5 $\pm$ 0.1 & 1397.3 $\pm$ 29.6 \\
              \cline{3-5}
              &           & 105-125 & 7.4 $\pm$ 0.1 & 1569.6 $\pm$ 16.8 \\
              \cline{3-5}
              &           & 125-145 & 8.5 $\pm$ 0.1 & 1912.3 $\pm$ 29.4 \\
              \cline{3-5}
              &           & 145-165 & 10.4 $\pm$ 0.2 & 2460.3 $\pm$ 40.9 \\
              \cline{3-5}
              &           & 165-185 & 14.7 $\pm$ 0.5 & 3635.4 $\pm$ 78.8 \\
              \cline{3-5}
              &           & 185-205 & 24.1 $\pm$ 1.3 & 5683.4 $\pm$ 199.6 \\
              \cline{3-5}
              &           & 205-225 & 52.6 $\pm$ 5.0 & 11408.9 $\pm$ 1016.4 \\
              \cline{3-5}
              &           & 225-245 & 146.1 $\pm$ 11.2 & 30377.8 $\pm$ 2204.0 \\
              \cline{3-5}
              &           & 245-265 & 262.6 $\pm$ 13.5 & 52793.7 $\pm$ 2625.3 \\
              \cline{3-5}
              &           & 265-285 & 441.5 $\pm$ 11.8 & 90199.4 $\pm$ 2151.5 \\
              \cline{3-5}
              &           & 285-300 & 205.8 $\pm$ 9.6 & 43187.2 $\pm$ 2028.7 \\
    \hline
        \multirow{9}{*}{-5.16}& \multirow{9}{*}{Holes} & 75-85 & 2.1 $\pm$ 0.1 & 291.6 $\pm$ 32.1 \\
              \cline{3-5}
              &           & 85-105 & 2.1 $\pm$ 0.1 & 303.4 $\pm$ 34.9 \\
              \cline{3-5}
              &           & 105-125 & 3.0 $\pm$ 0.1 & 505.0 $\pm$ 28.0 \\
              \cline{3-5}
              &           & 125-145 & 4.5 $\pm$ 0.2 & 912.5 $\pm$ 72.2 \\
              \cline{3-5}
              &           & 145-165 & 7.1 $\pm$ 0.3 & 1562.7 $\pm$ 83.7 \\
              \cline{3-5}
              &           & 165-185 & 12.7 $\pm$ 0.6 & 2868.2 $\pm$ 92.2 \\
              \cline{3-5}
              &           & 185-205 & 24.9 $\pm$ 1.9 & 5559.9 $\pm$ 342.2 \\
              \cline{3-5}
              &           & 205-225 & 65.8 $\pm$ 6.9 & 14317.0 $\pm$ 1478.4 \\
              \cline{3-5}
              &           & 225-245 & 186.9 $\pm$ 11.3 & 38594.3 $\pm$ 2098.6 \\
              \cline{3-5}
              &           & 245-265 & 301.1 $\pm$ 13.2 & 60410.5 $\pm$ 2563.7 \\
              \cline{3-5}
              &           & 265-285 & 492.5 $\pm$ 13.8 & 103303.9 $\pm$ 3409.4 \\
              \cline{3-5}
              &           & 285-300 & 158.7 $\pm$ 15.8 & 33897.7 $\pm$ 3268.8 \\
    \hline
    \end{tabular}}
    \caption{Summary of the mean peak amplitude and integrated pulse area across the temperature range probed for an electric field of $\pm$ 5.16 V/$\mu$m.}
    \label{tab:TempSummaryTable750V}
\end{table}

\clearpage

\newpage 

%%%%%%%%%%%%%%%%%%%%%%%%%%%%%%%%%%%%
\section{Modeling of the electric field}\label{app:EFieldModeling}
\begin{figure}[htb]
    \centering
    \includegraphics[scale=0.26]{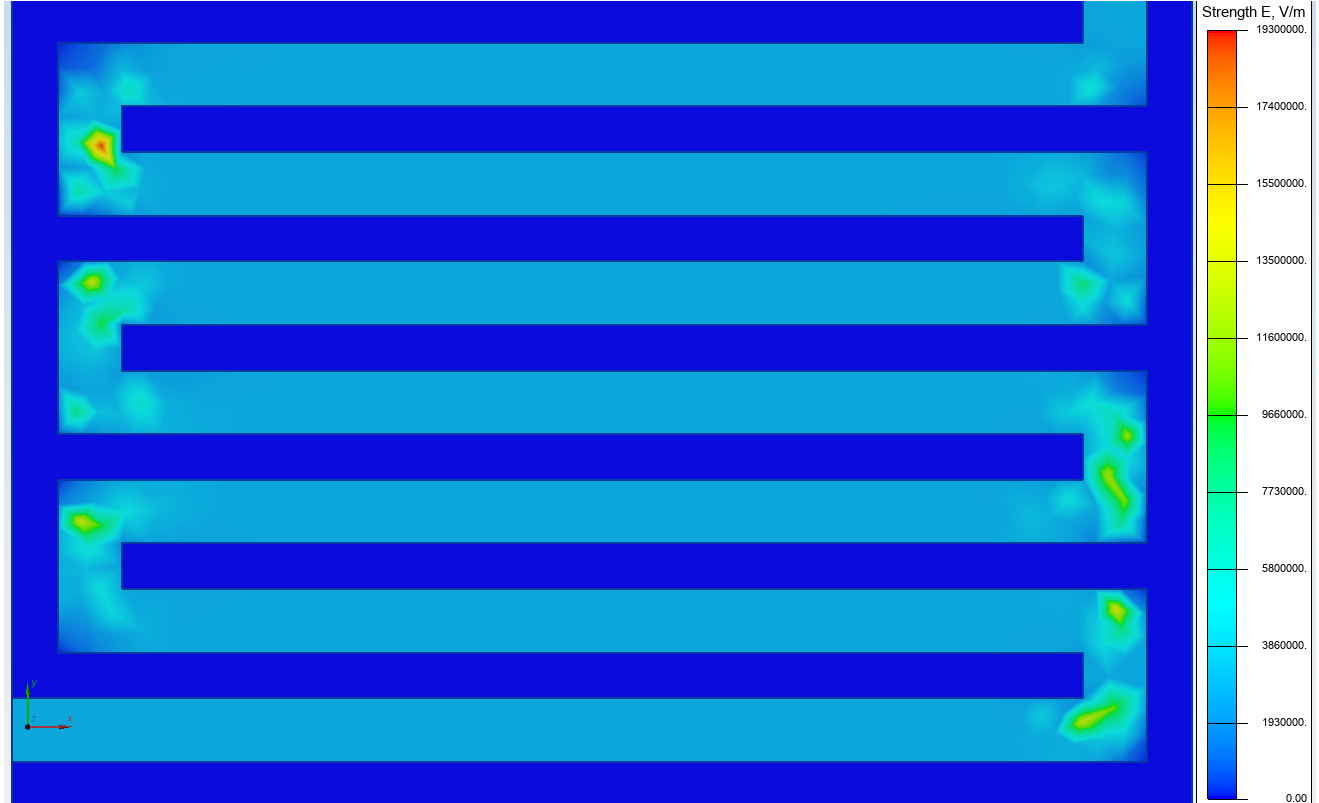}
    \includegraphics[scale=0.26]{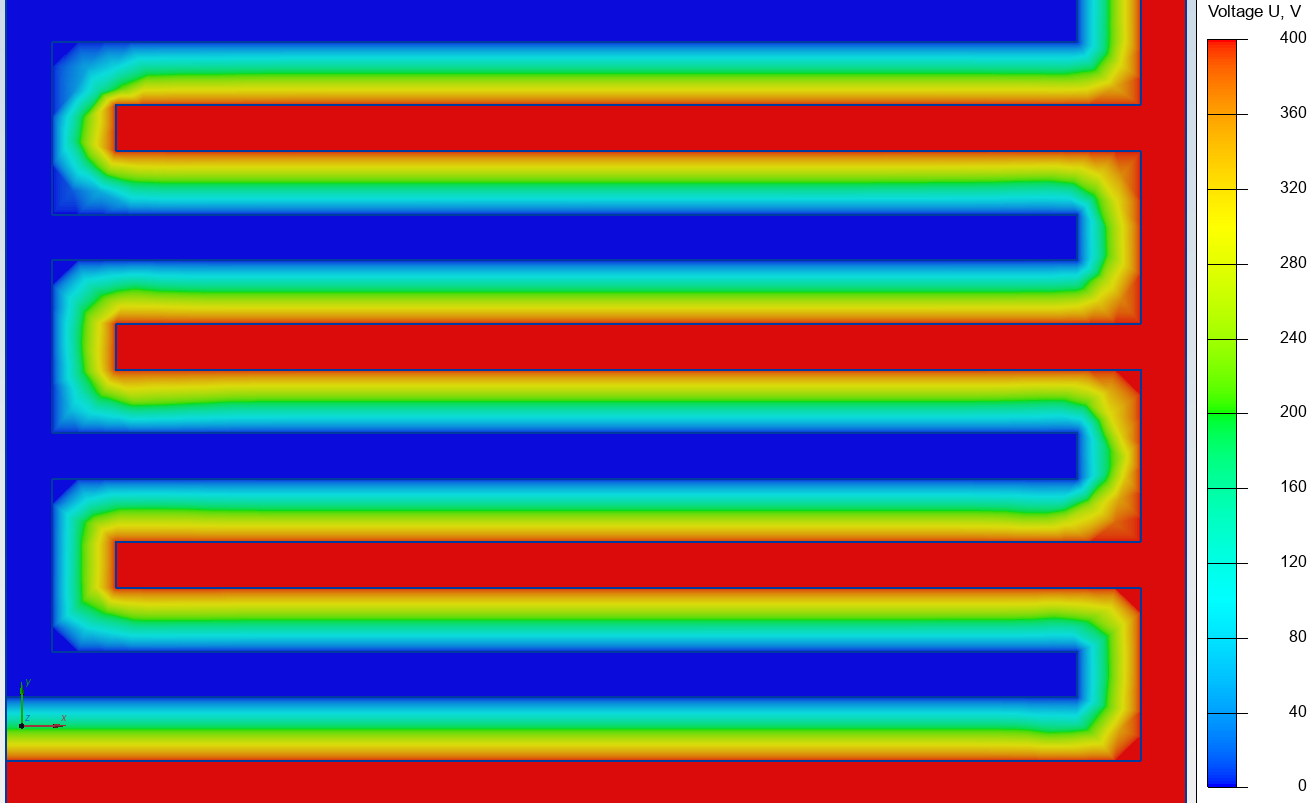}
    \includegraphics[scale=0.24]{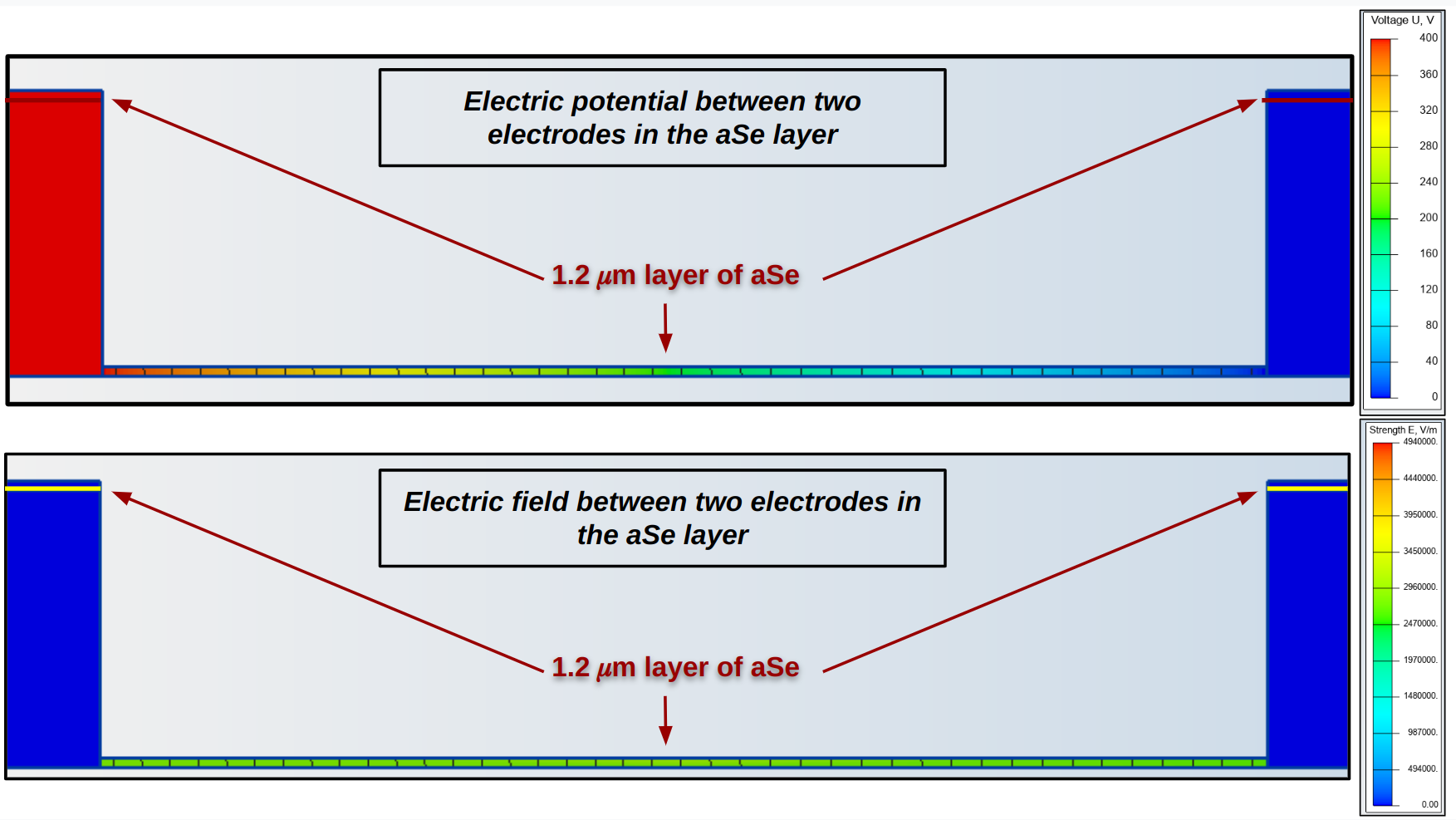}
    \caption{Top: The output of the QuickField simulation for the interdigitated PCB board used in this work from an overhead perspective showing the electric field (left) and electric potential (right) when a potential of 400 volts is applied. Bottom: The output of the QuickField simulation from a horizontal propspective where the height of the electrodes compared to the thickness of the selenium layer can be seen both showing the electric field and potential. }
    \label{fig:FieldModelOverview}
\end{figure}

%%%%%%%%%%%%%%%%%%%%%%%%%%%%%%%%%%%%%
Using the online 3D computer-aided design (CAD) design software Fusion360 \cite{autodesk_2022}, a model of the interdigitated electrode found in the `cookie' board was created with the appropriate spacings and materials. Traces are modeled as 107 $\mu$m (W) $\times 1123 \mu$m (L) $\times 35 \mu$m (H) with the gaps evenly distributed at 147 $\mu$m. The electrodes are assumed to be silver.  This model was then exported to a online electric field modeling tool, QuickField \cite{tera}, which was used to model the behavior of the electric field and electric potential in the presence of the aSe coating. The aSe layer is assumed to be 1.2$\mu$m thick and is present between the electrodes as well as on top of the electrodes. For simplicity, no aSe is found on the vertical walls of the electrodes. A permittivity of 5.8 and 6.9 is assumed for the aSe and Ag respecitvely.  An example output of the simulation for an applied potential of 400 Volts can be seen in Figure \ref{fig:FieldModelOverview}.

\begin{figure}[htb]
    \centering
    \includegraphics[scale=0.35]{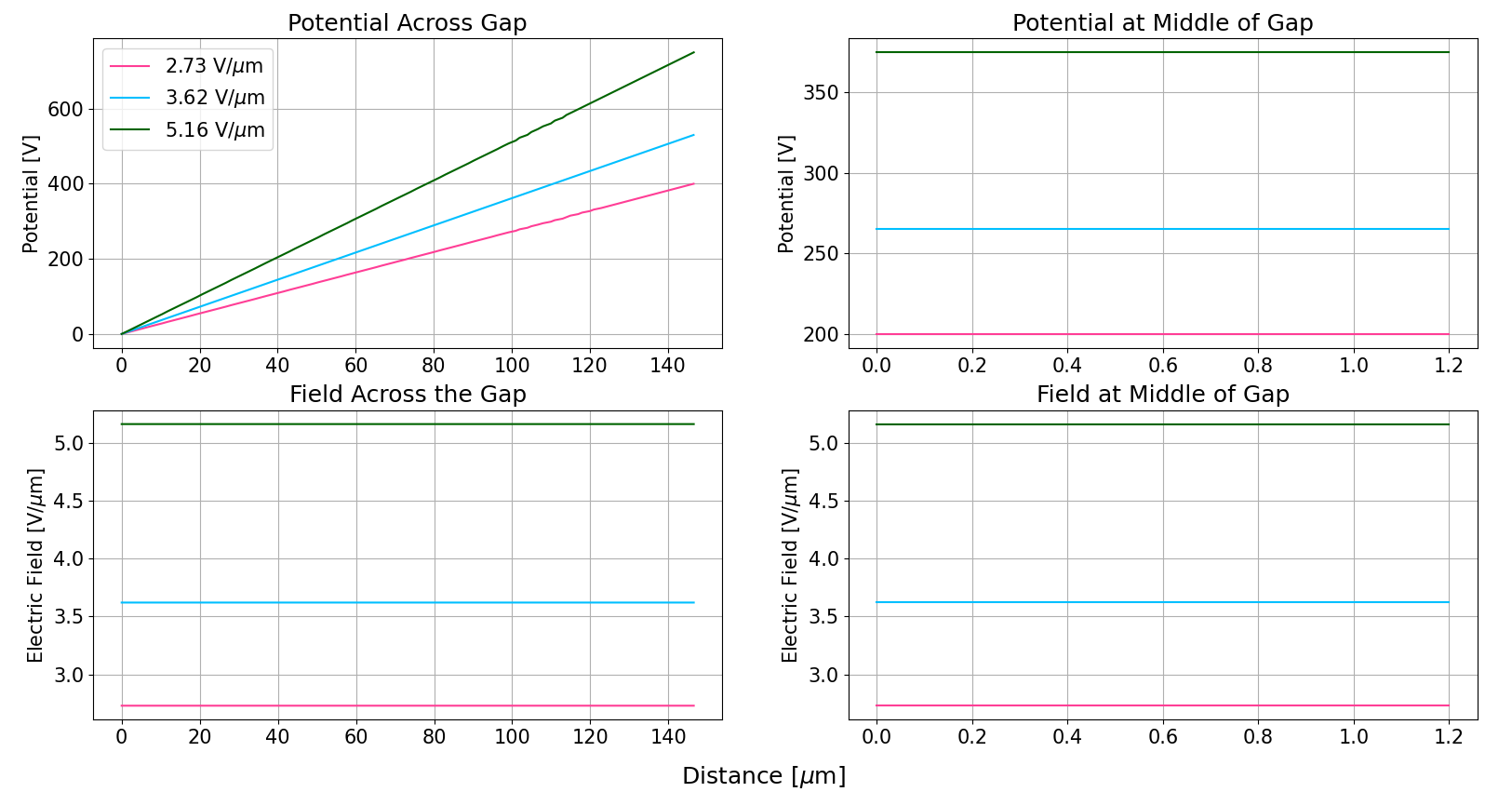}
    \caption{Left: The simulated electric potential (top) and electric field (bottom) for three different transport fields across the 147$\mu$m gap between electrodes in a region of high uniformity. The field can be seen at the expected value and uniform across the electrode while the potential changes linearly, as expected. Right: The simulated electric potential (top) and field (bottom) for a single position in the middle of the gap between the electrodes ($\sim 73.5 \mu$m from the electrodes) and then traversing the thickness of the aSe layer ($1.2 \mu$m). The electric field is found to be uniform as a function of the thickness of the selenium and the potential is a constant.}
    \label{fig:MiddleOfGap}
\end{figure}

Overall, the electric field (and the corresponding gradient in the electric potential) across the board and within the selenium is uniform and consistent with the estimated field calculated using the geometry of the board. This can be better seen in Figure \ref{fig:MiddleOfGap}, which shows analytically how simulation predicts the electric potential and field vary as you traverse the gap between electrodes (left) as well as picking a single point in the middle of the gap and moving vertically through the simulated layer of aSe (right). In both cases, the field and potential are found to be as expected.

\begin{figure}[htb]
    \centering
    \includegraphics[scale=0.35]{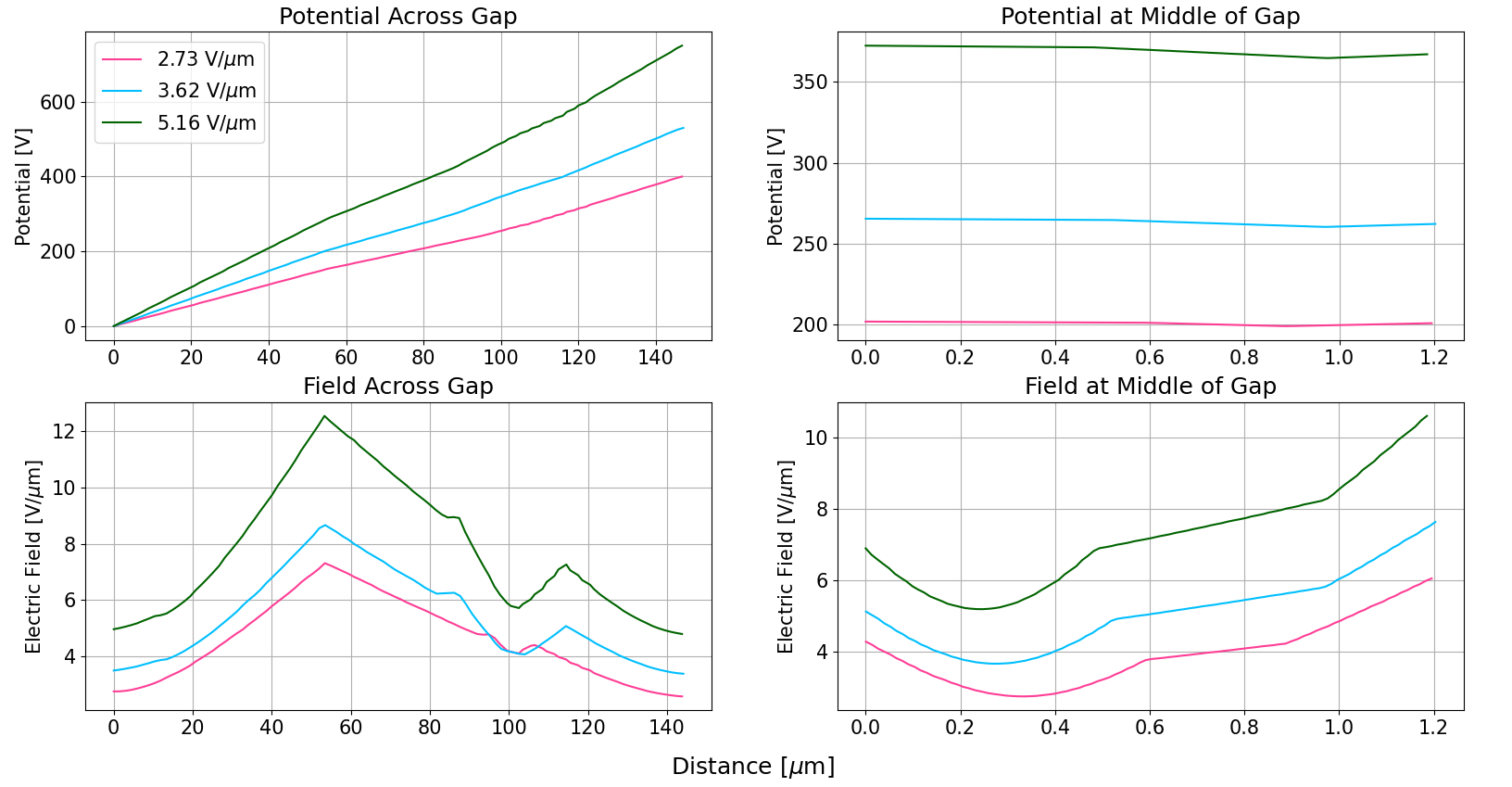}
    \caption{Left: The simulated electric potential (top) and electric field (bottom) for three different transport fields across the 147$\mu$m gap between electrodes in where the field was evidently non-uniform due to the edge effects of the electrode. The field can be seen to vary significantly from the geometrically calculated value due to edge and corner effects Right: The simulated electric potential (top) and field (bottom) for a single position in the middle of the gap between the electrodes ($\sim 73.5 \mu$m from the electrodes) and then traversing the thickness of the aSe layer ($1.2 \mu$m). Here too the electric field is found vary as a function of the thickness of the selenium.}
    \label{fig:NonUniformMiddleOfGap}
\end{figure}

The areas with the largest non-uniformity in the electric field occur in the regions where one electrode terminates in the gap of the opposite pair of interdigitated electrodes. In this region the edges of the electrodes cause the field to be more non-linear. Figure \ref{fig:NonUniformMiddleOfGap} shows analytically how the electric potential and field varies as you traverse the gap between electrodes (left) as well as picking a single point in the middle of the gap and moving vertically through the simulated layer of aSe (right). The point chosen here is an area where the variations can be seen in Figure \ref{fig:FieldModelOverview} to be the largest. Due to the geometric effects of the edges of the electrodes, the electric field can vary 2-3 times larger then the uniform region between the electrodes. While this is a large variation in the field, this effect is ultimately determined to be of little significance to the main analysis presented here. Since the non-uniform gap region represents a small fraction of the overall surface area of the board (the non-uniform area represents $< 1.5 \%$ of the total surface area), the effect on the reconstructed signal due to photons creating electron/hole pairs in this region is expected to be quite small.

%%%%%%%%%%%%%%%%%%%%%%%%%%%%%%%%%%%%%%%%%%%%%
%  The Repeatability here
%%%%%%%%%%%%%%%%%%%%%%%%%%%%%%%%%%%%%%%%%%%%%
\section{Repeatability}\label{app:repeat}

Table \ref{tab:TempSummaryTable400VRepeat} summarizes the values for the measurements of the peak amplitude and integrated pulse area taken during the main data taking campaign as well as the repeated measurements made multiple days later at $\pm$2.73 V/$\mu$m. The same analysis technique described in Section \ref{sec:DQandAnalysis} was utilized for the repeated measurement data set. The results are seen to vary between $\sim 10 - 100\%$ when the charge carriers are electrons and $\sim 10 - 300\%$ when the charge carriers are holes. The largest variation is seen in the integrated pulse area and upon inspection of the various waveforms, can be primarily attributed to a shift in the baseline and the subsequent calculation of the integrated area.  The room temperature measurements are seen to be generally consistent with one another, with the variation between tests being less than 25\% in both the peak amplitude and integrated area and the largest variations being seen at in the temperature bins between 205K and 105K.

\begin{table}[htb]
    \centering
    \resizebox{0.98\textwidth}{!}{%
    \begin{tabular}{|c|c|c|c|c|c|c|}
    \hline
        \textbf{Field} & \textbf{Charge Carrier} & \textbf{Temperature} & \textbf{Peak Amplitude} & \textbf{Repeat Measurement} & \textbf{Integrated Pulse Area} & \textbf{Repeat Measurement}  \\
        (V/$\mu$m) &  & (K) & (mV) & \textbf{Peak Amplitude} (mV) & (mV$\cdot \mu$s) & \textbf{Integrated Pulse Area} (mV$\cdot \mu$s) \\
    \hline
        \multirow{9}{*}{+2.73}& \multirow{9}{*}{Electrons} & 75-85 & 4.2 & 5.8 & 854.7 & 1173.5 \\
              \cline{3-7}
              &           & 85-105 & 4.4 & 5.6 & 890.1 & 1141.1 \\
              \cline{3-7}
              &           & 105-125 & 4.7 & 6.0 & 953.7 & 1319.0 \\
              \cline{3-7}
              &           & 125-145 & 5.0 & 7.1 & 1051.4 & 1586.6 \\
              \cline{3-7}
              &           & 145-165 & 5.4 & 9.2 & 1191.9 & 2162.1 \\
              \cline{3-7}
              &           & 165-185 & 6.7 & 12.9 & 1574.2 & 3046.1 \\
              \cline{3-7}
              &           & 185-205 & 10.1 & 21.1 & 2296.7 & 4630.1 \\
              \cline{3-7}
              &           & 205-225 & 22.5 & 35.1 & 4996.6 & 7360.0 \\
              \cline{3-7}
              &           & 225-245 & 72.4 & 76.8 & 15204.2 & 16572.6 \\
              \cline{3-7}
              &           & 245-265 & 119.2 & 144.2 & 23695.2 & 28149.9 \\
              \cline{3-7}
              &           & 265-285 & 179.4 & 246.0 & 36369.9 & 49208.5 \\
              \cline{3-7}
              &           & 285-300 & 75.2 & 84.4 & 15389.1 & 17107.3 \\
    \hline
        \multirow{9}{*}{-2.73}& \multirow{9}{*}{Holes} & 75-85 & 1.3 & 2.9 & 38.5 & 117.6 \\
              \cline{3-7}
              &           & 85-105 & 1.3 & 3.1 & 41.3 & 196.7 \\
              \cline{3-7}
              &           & 105-125 & 1.6 & 3.8 & 71.6 & 286.7 \\
              \cline{3-7}
              &           & 125-145 & 1.9 & 4.8 & 147.4 & 442.9 \\
              \cline{3-7}
              &           & 145-165 & 2.4 & 6.6 & 345.7 & 724.9 \\
              \cline{3-7}
              &           & 165-185 & 3.4 & 10.3 & 715.1 & 1521.0 \\
              \cline{3-7}
              &           & 185-205 & 6.1 & 16.9 & 1394.3 & 3622.4 \\
              \cline{3-7}
              &           & 205-225 & 16.8 & 31.2 & 3682.3 & 6698.1 \\
              \cline{3-7}
              &           & 225-245 & 51.0 & 59.5 & 10505.1 & 12175.4 \\
              \cline{3-7}
              &           & 245-265 & 71.6 & 86.0 & 14000.7 & 17374.9 \\
              \cline{3-7}
              &           & 265-285 & 87.7 & 103.2 & 17431.7 & 21528.1 \\
              \cline{3-7}
              &           & 285-300 & 46.4 & 52.7 & 9486.1 & 11136.7 \\
    \hline
    \end{tabular}}
    \caption{Summary of the peak amplitude and integrated pulse area found during the repeat measurement across the temperature range probed for an electric field of $\pm$2.73 V/$\mu$m.}
    \label{tab:TempSummaryTable400VRepeat}
\end{table}

\bibliographystyle{unsrt}
\bibliography{biblio}

\end{document}